\documentclass[10pt, journal]{IEEEtran}
\usepackage[babel]{csquotes}
\usepackage[backend=bibtex,style=ieee,natbib=true]{biblatex}

\usepackage{graphicx} 
\newcommand{\blue}[1]{\textcolor{black}{#1}}
\newcommand{\red}[1]{\textcolor{red}{#1}}
\usepackage{amsmath,amssymb,amsfonts}
\usepackage[symbol]{footmisc}
\usepackage{dsfont}
\usepackage{algorithm}
\usepackage{algpseudocode}
\usepackage{graphicx}
\usepackage{textcomp}
\usepackage{matlab-prettifier}
\usepackage{booktabs}
\usepackage{xcolor,colortbl}
\usepackage[caption=false]{subfig}
\usepackage{stfloats}
\usepackage{pbox}
\usepackage{makecell}
\usepackage{amsmath}
\usepackage{steinmetz}
\usepackage{tabularx}
\usepackage{multirow}
\usepackage{adjustbox}

\usepackage{tikz}
\usetikzlibrary{arrows.meta}
\usetikzlibrary{%
    decorations.pathreplacing,%
    decorations.pathmorphing%
}
\usepackage{hyperref}
\hypersetup{
    colorlinks,
    linkcolor={red!50!black},
    citecolor={blue!50!black},
    urlcolor={blue!80!black}
}
\usepackage{verbatim}
\usepackage[american,cuteinductors,smartlabels]{circuitikz}
\usetikzlibrary{math}
\usepackage{amsmath}
\tikzset{%
    body/.style={inner sep=0pt,outer sep=0pt,shape=rectangle,draw,thick,pattern=north east lines wide},
    dimen/.style={<->,>=latex,thin,every rectangle node/.style={fill=white,midway,font=\sffamily}},
    symmetry/.style={dashed,thin},
}

\begin{document}
% \title{Loewner Matrix Based Fast Frequency Sweep Method for Electromagnetic Modeling of Multiport Systems }
\title{Fully-Adaptive and Semi-Adaptive Frequency Sweep Algorithm Exploiting Loewner-State Model for EM Simulation of  Multiport Systems}
\author{Shilpa T. N.\textsuperscript{\href{https://orcid.org/0009-0008-5669-6138}{\includegraphics[scale=0.1]{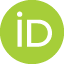}}} and Rakesh~Sinha\textsuperscript{\href{https://orcid.org/0000-0003-0592-8505}{\includegraphics[scale=0.1]{fig_new/orcid.png}}},~\IEEEmembership{Member,~IEEE}
        % <-this % stops a space
\thanks{\red{This work has been accepted by the IEEE Transactions on Microwave Theory and Techniques (\url{https://doi.org/10.1109/TMTT.2025.3557208}) for possible publication. Copyright
may be transferred without notice, after which this version may no longer be
accessible.}  This article has supplementary downloadable material available at
\url{http://ieeexplore.ieee.org}, provided by the authors. This includes additional results generated using the proposed method. (\textit{Corresponding author: Rakesh Sinha.})}

\thanks{Authors are with the Department of Electrical Engineering, National Institute of Technology Rourkela, Rourkela-769008, India (e-mail: r.sinha30@gmail.com, shilpatjayaprakash@gmail.com)}
\thanks{Color versions of one or more of the figures in this paper are available online at http://ieeexplore.ieee.org.}}
\markboth{IEEE Transactions on Microwave Theory and Techniques,}%
{T N and Sinha: Loewner Matrix Based Fast Frequency Sweep Method for Electromagnetic Modeling of Multiport Systems}
\maketitle
\begin{abstract}

This paper employs a fully adaptive and semi-adaptive frequency sweep algorithm using the Loewner matrix-based state model for the electromagnetic simulation. The proposed algorithms use two Loewner matrix models with different or the same orders with small frequency perturbation for adaptive frequency selection. The error between the two models is calculated in each iteration, and the next frequency points are selected to minimize maximum error. With the help of memory, the algorithm terminates when the error between the model and the simulation result is reached within the specified error tolerance. In the fully adaptive frequency sweep algorithm, the method starts with the minimum and maximum frequency of simulation.  In the semi-adaptive algorithm,  a novel approach has been proposed to determine the initial number of frequency points necessary for system interpolation based on the electrical size of the structure. The proposed algorithms have been compared with the Stoer–Bulirsch algorithm and Pradovera's minimal sampling algorithm for electromagnetic simulation. Four examples are presented using MATLAB R2024b. The results show that the proposed methods offer better performance in terms of speed, accuracy and the requirement of the minimum number of frequency samples. The proposed method shows remarkable consistency with full-wave simulation data, and the algorithm can be effectively applicable to electromagnetic simulations.

\end{abstract}

\vspace*{-1em}

\section{Introduction}
Mathematical modelling \cite{MBPE1}, \cite{MBPE2}, \cite{MBPE3}, \cite{Burke_MBPE_1989}, \cite{AFS1991} of system response or system function (transfer function) plays an important role in electrical and electronics engineering in understanding the system behaviour for given input or excitation. One of the most popular models is rational polynomial approximation \cite{Burke_MBPE_1989} of the transfer function. Both applied mathematicians \cite{Barry1986}, \cite{antoulas1986scalar}, \cite{MAYO2007634}, \cite{AAA2018} and engineers \cite{Burke_MBPE_1989}, \cite{Gustavsen_VF99}, \cite{S.Lefteriu_Loewner2010} played important roles in the development of the rational polynomial model of functions using interpolation techniques. Some of the well-known models are rational polynomial \cite{Burke_MBPE_1989}, sum of exponential \cite{Matrix_pencil_TKS95}, \cite{Keyhan_Matrix_Pencil2012}, pole-residue expansion (partial fraction expansion) \cite{VF_Gustavsen99}, state-space \cite{MAYO2007634}, \cite{S.Lefteriu_Loewner2010}, barycentric polynomial \cite{Barry1986}, continued fraction  \cite{AFS2001},  etc.

Various techniques have been developed to fit the data of electromagnetic and circuit parameters in mathematical models. The most popular techniques are vector fitting (VF) \cite{VF_Gustavsen99}, matrix pencil \cite{Matrix_pencil_TKS95}, \cite{Keyhan_Matrix_Pencil2012}, asymptotic waveform evaluation \cite{AWE1990}, Stoer–Bulirsch Algorithm \cite{AFS2-2003}, \cite{AFS4-2013}, Cauchy's Technique \cite{Kottapalli1991}, \cite{AFSTKS1997}, \cite{AFS1998}, least square fitting, Frequency Domain Prony Method \cite{FD_Prony2020}, Loewner Matrix Method \cite{MAYO2007634} \cite{S.Lefteriu_Loewner2010}, AAA (Adaptive Antoulas-Anderson) algorithm \cite{AAA2018}, \cite{FastAAA2017}, \cite{StablePoleAAA2021}, \cite{Multi_function_AAA2020}, Lagrange’s interpolation \cite{Zhu}, Thiele Interpolation \cite{AFS2001} etc.

For electrically large electromagnetic structures or higher-order electric circuits, it takes a significant amount of time to simulate the circuit or geometry at each frequency sample. In broadband simulation, the number of frequency samples ($N$) increases depending on the frequency range ($f_{max}-f_{min}$) and resolution $\Delta f$ (difference between two successive frequency samples). Modelling the simulation (electromagnetic or circuit) results using a few frequency samples $n$ and interpolating the results to $N$ frequency points is the subject of fast frequency sweep (FFS). If the $n$ frequency points are selected adaptively to reduce the error between the actual result $H(s)$ and predicted model $\widetilde{H}(s)$, the method will be called adaptive frequency sweep (AFS). Please note that all AFS can be called FFS, but not all FFS are necessarily adaptive in nature or AFS.  

Most of the commercial EM simulators use the asymptotic waveform evaluation (AWE) \cite{AWE1990}, \cite{CFH1995}, \cite{chiprout1994asymptotic}, \cite{cockrell1996asymptotic}, \cite{AWE1998}, \cite{Jeong_AWE2016} algorithm for AFS. The AWE was initially developed for time domain analysis of electrical circuits \cite{AWE1990}. Later, it was applied for the Method of Moments (MoM) and finite element method (FEM) analysis of electromagnetic scattering \cite{AWE1998} and microwave circuit \cite{Dan_Jiao_AWE99}, \cite{Jeong_AWE2016} analysis. In the AWE technique, a Taylor series expansion/ Pade approximation is applied to the matrix equation (of the MoM or FEM) around a specific frequency. The coefficients of the Taylor series/ Pade are derived from the frequency derivatives of the matrices evaluated at the expansion frequency. These coefficients are then utilized to estimate the system's frequency response across a specified frequency range. In asymptotic waveform evaluation, explicit moment matching was performed to obtain the dominating poles using the Pade approximation. However, the AWE approach is mathematically unstable when approximating higher-order moments. The equivalent circuit-based technique is a hybrid process; it first solves the EM portion of the system using fundamental methods such as the Method of Moments (MoM) or the Finite Element Method (FEM) and then approximates the circuit portion using Modified nodal analysis or other circuit approximation \cite{Ruheli_Equivalent_ckt_1974}.  However, this method has a downside, which becomes clear when dealing with RF devices with many ports. In such cases, the equivalent circuit model may become rather complex, and calculating and extracting the characteristic parameters for each element inside the model can be time-consuming \cite{Li_firstorder_2021}.

One of the early works on rational polynomial-based AFS  was presented in \cite{AFS1992}, where segmented rational models of the EM transfer function were obtained using a few samples in each segment. Rational polynomial base AFS has recently been explored in \cite{AFS2019}, where interpolating and testing frequency points are selected using the successive arithmetic mean of adjacent frequency points. The use of local rational interpolation (or segmented rational interpolation) has been demonstrated in \cite{LRM2021} for modelling measurement results and in \cite{AFS-2024} for adaptive frequency sampling. 

A continued fraction rational polynomial based Stoer–Bulirsch (S–B) algorithm \cite[sec. 2.2]{stoer1980introduction} has been used in \cite{AFS2-2003}, \cite{AFS4-2013} for AFS. The adaptive frequency points were selected to minimize the error between two competitive Thiele-type rational polynomials. The vector fitting \cite{VF_Gustavsen99} algorithm-based AFS algorithm was proposed in \cite{AFS3-2008}, where adaptive frequency points were selected to minimize the error between two competitive pole-residue type rational polynomials. The segmented non-rational Lagrange interpolation and Spline Interpolation based AFS proposed in  \cite{Zhu} and \cite{MartinezSpline2019}. Scalar rational polynomial-based AFS techniques presented in \cite{AFS1992},  \cite{AFS2019}, \cite{AFS-2024}, \cite{AFS2-2003}, \cite{AFS3-2008} may not be suitable for large-scale multiport networks. As the speed-up factors of rational interpolation-based AFS algorithms have not been reported, it is difficult to say which algorithms work better in terms of speed-up factor.   

Recently, Hongliang Li \textit{et al.} \cite{Li_firstorder_2021}  introduced an FFS method for miniature passive RF circuits, exploiting the Analytic Extension of Eigenvalues (AEE). This technique uses eigenvalue decomposition on the Z-parameters of circuit components at a single or a few frequencies and employs analytic extension to compute eigenvalues across the entire frequency range of interest. Their findings highlight the high accuracy of the proposed AEE approach when applied to miniature RF circuits with electrical sizes less than one-tenth of the wavelength. Furthermore, it demonstrates a broader and more robust applicability than methods reliant on lumped equivalent circuits. Hongliang Li \textit{et al.}  \cite{Second-Order} updated the AEE method \cite{Li_firstorder_2021} by a second-order analytic extension of eigenvalue, and this method applied to miniaturised RF circuits whose electrical sizes are up to one wavelength. 

Curve-fitting methods used in frequency domain admittance characterization \cite{SALARIEH2021107254}, \cite{GURRALA2021107345}, \cite{Morales_FDF_2020} can have potential applications in fast frequency sweep of s-parameters data obtained using full-wave EM simulation. The commonly used curve-fitting techniques in power-system applications are the vector-fitting \cite{VF_Gustavsen99}, matrix pencil method \cite{Matrix_pencil_TKS95}, \cite{Keyhan_Matrix_Pencil2012}, and the Loewner matrix method \cite{Kabir_Loewner_2012}. Among these techniques, the Loewner matrix method can be most suitable for multiport s-matrix characterization over a wide band of frequency.    

The introduction of the vector fitting algorithm marked a significant advancement in the development of macromodeling tools and algorithms \cite{Gustavsen_VF99}. The core of the vector fitting approach is fitting a rational function model to a set of tabular data in the frequency domain or time domain. Using iterative processes, VF finds the solution and generates a pole-residue representation of the model, which can be easily converted into the state space form. When trying to fit high-order rational functions or working with massive data sets, the processing requirements of vector fitting may need to be revised. Choosing the right rational function order may be challenging when using the vector fitting approach \cite{S.Lefteriu_Loewner2010}. The VF also face convergence issues \cite{Convergency_VF2013}, \cite{Convergency_VF2016}. 

A Loewner-Matrix (LM) based FFS algorithm of high-speed modules is proposed in \cite{Kabir_FFS1_2012}. The most significant drop in the singular value plot of the Loewner matrix pencil is chosen for the system's order and termination condition of FFS. New frequency sample points are chosen to maintain uniform sampling, which may not provide optimum accuracy. In the case of a distributed system, the drop may not be as significant as that of the lumped system, so \citeauthor{kabir2} modified the algorithm \cite{kabir2}. Determining model order was accomplished iteratively by considering the index of any of the five most substantial drops. The indices corresponding to the first five significant drops were arranged in descending order based on the drop values. The observation proposed in \cite{kabir2} may not be valid for arbitrary distributed multiport networks. Another Loewner-Matrix-based AFS has been presented in \cite{Lucas24}. The method is based on the surrogate model proposed in \cite[eq. (29)-(34)]{S.Lefteriu_Loewner2010} by \citeauthor{S.Lefteriu_Loewner2010}. The adaptive frequency points are selected based on the pseudo-error between \cite[eq. (33)]{S.Lefteriu_Loewner2010} and \cite[eq. (34)]{S.Lefteriu_Loewner2010}. However, the technique involves two matrix inverse/ solving problems for each model, and evaluation of the model for each iteration may take a good amount of time.     

Recently, \citeauthor{greedy}  in \cite{greedy} proposed a greedy barycentric interpolation technique based on the null space of the Loewner matrix. The Loewner matrix was obtained using positive-negative partitioning of sample data. The algorithm proposed in \cite{greedy} is a modification of AAA (adaptive Antoulas–
Anderson) algorithm \cite{AAA2018}. The barycentric representation of the rational function and adaptive frequency point selection are the foundations of the AAA algorithm. In the AAA algorithm, the next frequency point is chosen based on the error between the function value at the testing frequency points and the rational interpolation values at all testing points; the point where the error is maximum is chosen as the next frequency point. The AAA approach cannot directly be employed for FFS since we need already simulated values at testing frequency points, which will be time-consuming. In \cite{greedy}, the adaptive frequency point is selected based on pseudo error, which is inversely proportional to the denominator of the rational rational approximation. The algorithm proposed in \cite{greedy} (Pradovera's algorithm) is one of the suitable candidates for AFS. In this work, we will investigate the effectiveness of Pradovera's algorithm in AFS applications. 

One of the problems with adaptive frequency sweep is that it requires evaluation of the system model at all frequencies at each iteration for different numbers of input samples until the model converges to actual EM simulation. The evaluation of the approximation model takes some time. Even though the model evaluation time is less than the EM simulation time, it becomes significant for large systems with slow convergence. We can reduce the system evaluation time by making good guesses about the system order. Guessing the system order is tricky, as larger estimation may lead to over-estimation of system order, while smaller estimation leads to slow convergence. In this work, we have devised a technique to estimate the system order based on the total electrical lengths of PCB traces. It has been shown that there exists a third-order equivalent circuit model of 0.2$\lambda$ transmission line (TL). If we consider that all PCB traces are cascade connections of small 0.2$\lambda$, then we can estimate the system order from the electrical length at the highest frequency. A direct use of the initial system order may not give the desired accuracy in FFS. However, we can add new samples adaptively to reach the desired accuracy. New frequency points have been added based on the pseudo error between two LM-state models with frequency perturbation and order reduction. The adaptive frequency sweep can be terminated based on actual error in the new sample. This frequency sweep method will be referred to as the semi-adaptive frequency sweep, as initial samples are collected using a non-adaptive process. The proposed algorithm can also used in fully adaptive frequency configurations with minimum and maximum frequency as initial samples. It has been shown here that an optimum logarithmic initial frequency sampling can provide better convergence than uniform sampling.      The proposed algorithms have been compared against Pradovera's algorithm \cite{greedy} and Stoer-Bulirsch path-II algorithm \cite{simo:24}. Multiple implementations were showcased using MATLAB RF PCB Toolbox, and it shows the proposed method offers better performance with a minimum number of frequency points.   

% In addition to that, we have introduced a controlled logarithmic sampling technique, which provides better accuracy compared to uniform sampling. We explored various data partitioning and sampling techniques to identify the most effective strategies for precise predictions. Also, we have compared our method with Pradovera's algorithm \cite{greedy}. 

The rest of the manuscript is organized as follows. Initially, Loewner matrix-based system approximation theory for SISO and MIMO systems has been presented in Section \ref{sec:theory}. The procedure of the proposed semi-adaptive and fully adaptive LM-based frequency sweep algorithm has been explained in Section \ref{sec:salm}. At first theory behind initial sample collection has been discussed in Section \ref{sec:order} and \ref{sec:sampling}. The proposed partitioning scheme for Loewener matrix formation has been discussed in Section \ref{sec:partition}. The reduction of redundant order of LM state matrices has been explained in Section \ref{sec:mor}. The most important step: adaptive frequency selection is elaborated in Section \ref{sec:adaptive}, and the complete algorithm is also showcased.     Four different examples are presented to demonstrate the effectiveness of the proposed algorithms in Section \ref{sec:RESULTS}. Finally, the paper is concluded in Section \ref{sec:Conclusion}. 

% Section \ref{sec:prado} explains the theory behind Pradovera's algorithm. One of the most stable AFS algorithms based on the Stoer-Bulirsch Interpolating function has been presented in Section \ref{sec:SB}.
\section{Loewner Matrix based Approximation of EM Simulation Results}\label{sec:theory}

\subsection{State-space form representation of Transfer Function using Loewner Framework}
In the electromagnetic simulation, the port-to-port interaction of multiport networks can be represented using network parameters. The scattering parameters representation is commonly used to represent multiport circuits' input/ output behaviour. Other parameters like impedance (Z) or admittance (Y) can be obtained from the S-parameters. Throughout this paper, all the network parameters will be considered as transfer function (TF) $H\in \{S, Y, Z\}$. It has been assumed that $H$ is a continuous function of complex frequency $s=j\omega$ (i.e., $H(s)$). Therefore, the transfer function $H$ can be approximated in terms of $n^{th}$ order state matrices as \cite{S.Lefteriu_Loewner2010}

\begin{align}
    \widetilde{H}_n(s)=\textbf{C}(s\textbf{E}-\textbf{A})^{-1}\textbf{B}+D\quad \text{for} \quad s_{min}\leq s\leq s_{max}
    \label{eq:Hs1}
\end{align}
for a single-input and single-output (SISO) system. Where $s_{min}=j2\pi f_{min}$  and $s_{max}=j2\pi f_{max}$ are minimum and maximum frequency samples used for the approximation. The representation in \eqref{eq:Hs1} has been obtained from the $n^{th}$ order descriptor system representation of the linear time-invariant (LTI) system \cite{S.Lefteriu_Loewner2010}:
\begin{subequations}
    \begin{align}
        \textbf{E}\dot{\textbf{x}}&=\textbf{A}\textbf{x}+\textbf{B}u\\
        y&=\textbf{C}\textbf{x}+Du
    \end{align}
\end{subequations}
where $\textbf{x}\in \mathbb{C}^{n\times1}$  is state vector, $u\in \mathbb{C}^{1\times1}$ and $y\in \mathbb{C}^{1\times1}$ are input and output signals. $\textbf{A}\in \mathbb{C}^{n\times n}$, $\textbf{E}\in \mathbb{C}^{n\times n}$, $\textbf{B}\in \mathbb{C}^{n\times 1}$, $\textbf{C}\in \mathbb{C}^{1\times n}$ and $D\in \mathbb{C}^{1\times 1}$ are state matrices. We can consider $D=0$ for ease of state modelling purposes, and TF in \eqref{eq:Hs1} can be written as 
\begin{align}
    \widetilde{H}_n(s)=\textbf{C}(s\textbf{E}-\textbf{A})^{-1}\textbf{B}.
    \label{eq:Hs2}
\end{align}
The approximate model  $\widetilde{H}_n(s)$ is obtained from the actual TF $H(s)$ sampled at $n$ frequency points with $\widetilde{H}_n(s_j)=H(s_j)$ for $j\in [\![1, n ]\!]$. From $n$ sample data points of a given TF, we can predict the $n^{th}$ order state model representation of the TF $H(s)$ as given in \eqref{eq:Hs2}. 

Consider that a one-dimensional transfer function $H(j\omega)$ or $H(s)$ is discretised at $n$ discrete frequency points $[\![\omega_1,\omega_n]\!]$. The sampled data of the transfer function can be written as a vector $[H_1, H_2,..., H_n]$. As the data is generated from the LTI system, it has a conjugate symmetric property with respect to the zero frequency axis. Based on the conjugate symmetry property, we can expand the data set to 2$n$ number of sample with complex frequency set $\textbf{s}=[-j\omega_n,-j\omega_{n-1},...,-j\omega_2,-j\omega_1,j\omega_1,j\omega_2,...,j\omega_{n-1},j\omega_n]$ and transfer function set $\textbf{H}=[\bar{H}_n, \bar{H}_{n-1},..., \bar{H}_2, \bar{H}_1, H_1, H_2,..., H_{n-1}, H_n]$. We can partition the data into two sets $\{\textbf{s}_a, \textbf{H}_a\}$ and $\{\textbf{s}_b, \textbf{H}_b\}$, where $\textbf{s}_a=[s_{a1}, s_{a2},..., s_{an}]$ and $\textbf{H}_a=[H_{a1}, H_{a2},..., H_{an}]$ are $1\times n$ row vectors and $\textbf{s}_b=[s_{b1}, s_{b2},..., s_{bn}]^T$ and $\textbf{H}_b=[H_{b1}, H_{b2},..., H_{bn}]^T$ are $n \times 1$ column vectors. Then, the general expression of the state-matrices can be obtained as 

\begin{align}
\begin{split}
   \textbf{E}=-\textbf{W}_b\mathbb{L}\textbf{W}_a; \hspace{1em}& \textbf{A}=-\textbf{W}_b\sigma\mathbb{L}\textbf{W}_a;\\
    \textbf{B}=\textbf{W}_b\textbf{H}_b;\hspace{1em} & \textbf{C}=\textbf{H}_a\textbf{W}_a. \label{eq:EABCg}
\end{split}
\end{align}

where $\mathbb{L}$ and $\sigma\mathbb{L}$ are Loewner and shifted-Loewner matrices as defined in \eqref{eq:Lij}. $\textbf{W}_a\in\mathbb{C}^{n\times n}$ and $\textbf{W}_b\in\mathbb{C}^{n\times n}$ are random complex matrices of rank $n$. The proof of \eqref{eq:EABCg} has been discussed in Appendix-\ref{sec:proof}. Please note that we will consider $\textbf{W}_a=\textbf{W}_b=\mathds{1}$ in \eqref{eq:EABCg} throughout the manuscript, which leads to \cite{S.Lefteriu_Loewner2010}

\begin{align}
    \textbf{E}=-\mathbb{L}; \hspace{1em}& \textbf{A}=-\sigma\mathbb{L}; \hspace{1em} &
    \textbf{B}=\textbf{H}_b;\hspace{1em} & \textbf{C}=\textbf{H}_a. \label{eq:EABC}
\end{align} 

This has been considered, as the value of $H(s)$ is independent of $\textbf{W}_a$ and $\textbf{W}_b$. It is interesting to know that with proper choices of $\textbf{W}_a$ and $\textbf{W}_b$, we can make the system matrices $\textbf{E}$, $\textbf{A}$, \textbf{B} and \textbf{C} real.   Please note that the accuracy of the model depends upon the choice of $n$ and choices of sample data locations over the frequency span.

\subsection{State-representation of MIMO system}
		We must provide a specific interpolation strategy to ascertain the system's evolution to implement the Loewner framework. In the context of Loewner matrix interpolation, two established interpolation strategies are Vector Format Tangential  Interpolation (VFTI) \cite{S.Lefteriu_Loewner2010} and Matrix Format Tangential Interpolation (MFTI) \cite{Wang_MFTI_2010}, \cite{GURRALA2021107345}. VFTI will not work for a large number of ports since it will take more data for interpolation compared to MFTI for the same accuracy. The primary advantage of MFTI over VFTI is its ability to leverage a large amount of information inherent in the sampled matrices. Consequently, MFTI necessitates fewer samples to reconstruct the system, leading to enhanced accuracy when interpolating under-sampled, noisy, or ill-conditioned data. For a $p$ port network, if we are taking $n$ number of frequency points, the order of $\mathbb{L}$ and  $\sigma\mathbb{L}$ matrix in case of VFTI will be $\textit{n} \times \textit{n} $. For MFTI, the order of the  $\mathbb{L}$  and  $\sigma\mathbb{L}$ matrices will be $ pn  \times pn$,  so the MFTI will capture more data to predict the system accurately. In this work, we will use a vectorized implementation of the Loewner matrix for MFTI. For 1-dimensional data, if we are taking $n$-number of samples, all the corresponding matrices will have a size of $n\times n $ in MFTI. 
        % For a $p$-port network, the S matrix contains $ p \times p $ block elements. 

\subsubsection{Vector Format Tangential Interpolation}
 When the system is single-input and single-output (SISO), we can directly use the Loewner matrix framework as given in \eqref{eq:EABC}. On the other hand, if the system is multiple input multiple output (MIMO), we can not directly use the Loewner matrix framework as given in \eqref{eq:EABC}. However, with the help of tangential interpolation, we can predict the state of the system. The transfer function matrices are projected using right and left tangential directions. The right tangential direction and data matrices are given as  
 \begin{align*}
 \textbf{R} = [\textbf{r}_1, ..., \textbf{r}_n]\in \mathbb{C}^{p\times n}\\
 \textbf{W} = [\textbf{w}_1, ..., \textbf{w}_n]\in \mathbb{C}^{p\times n}
\end{align*}
where $\textbf{r}_i\in \mathbb{C}^{p\times 1}$ and $\textbf{w}_i\in \mathbb{C}^{p\times 1}$ are right tangential direction and data vector, respectively, with $i\in 1, ..., n$. The right tangential data vectors are obtained from right transfer function matrices $\textbf{H}_{ai}\in \mathbb{C}^{p\times p}$ using
\begin{align*}
\textbf{w}_i=\textbf{H}_{ai}\textbf{r}_i.
\end{align*}
The simplest choice of $\textbf{r}_i$ is the column vector of the identity matrix of order $p$. We can write $\textbf{r}_i=\textbf{e}_m \in \mathbb{R}^{p\times 1}$, with $m=((i-1)\bmod p)+1$. Where $\textbf{e}_m=\mathds{1}[:,m]$ is the $m^{th}$ column of identity matrix $\mathds{1}\in\mathbb{R}^{p\times p}$. 
The left tangential direction and data are given as 
\begin{align*}
\textbf{L}=\begin{bmatrix}
\textbf{l}_1\\
\vdots\\
\textbf{l}_n
\end{bmatrix}=[\textbf{R}]^{T}\in  \mathbb{C}^{n\times p}, \hspace*{2em}
\textbf{V}=\begin{bmatrix}
\textbf{v}_1\\
\vdots\\
\textbf{v}_n
\end{bmatrix}\in \mathbb{C}^{n\times p}
\end{align*}
where $\textbf{l}_i=\textbf{r}_i^T\in \mathbb{C}^{1\times p}$ and $\textbf{v}_i\in \mathbb{C}^{1\times p}$ are the left tangential direction and data vector. The left tangential data vectors are obtained from right transfer function matrices $\textbf{H}_{bi}\in \mathbb{C}^{p\times p}$ using
\begin{align*}
\textbf{v}_i=\textbf{l}_i\textbf{H}_{bi}.
\end{align*}
The $\textbf{l}_i$ can be chosen in same way as of $\textbf{r}_i$, with $\textbf{l}_i=\textbf{e}_m^T \in \mathbb{R}^{1\times p}$, with $m=((i-1)\bmod p)+1$.

Using the tangential interpolation, we can write a state representation of the MIMO system as 
\begin{subequations}
\begin{align}
 \textbf{E}=-\mathbb{L}\in \mathbb{C}^{n\times n}\\
\textbf{A}=-\sigma\mathbb{L}\in \mathbb{C}^{n\times n}\\
 \textbf{B}=\textbf{V}\in \mathbb{C}^{n\times p}\\
 \textbf{C}=\textbf{W}\in \mathbb{C}^{p\times n}
\end{align}
\label{eq:loewner_state1}
\end{subequations}
where, the elements of $\mathbb{L}$ and $\sigma\mathbb{L}$  can be represented as 
\begin{subequations}
\begin{align}
	\mathbb{L}_{ij}=\frac{\textbf{l}_i\textbf{H}_{aj}\textbf{r}_j-\textbf{l}_i\textbf{H}_{bi}\textbf{r}_j}{s_{aj}-s_{bi}}\\
	\sigma \mathbb{L}_{ij}=\frac{s_{aj}\textbf{l}_i\textbf{H}_{aj}\textbf{r}_j-s_{bi}\textbf{l}_i\textbf{H}_{bi}\textbf{r}_j}{s_{aj}-s_{bi}}  
\end{align}
\end{subequations}  

The limitation of tangential interpolation is that we predict $n^{th}$ order system from $p\times p\times n$ data points. If $p$ is large and $n$ is small, it leads to underestimation of the system. To solve this problem, we can utilize MFTI, which estimates ${np}^{th}$ order system from $p\times p\times n$ data points.  
\begin{figure}[!h]

\vspace*{-1.0em}
    \centering
  \centering
\includegraphics[width= 7 cm]{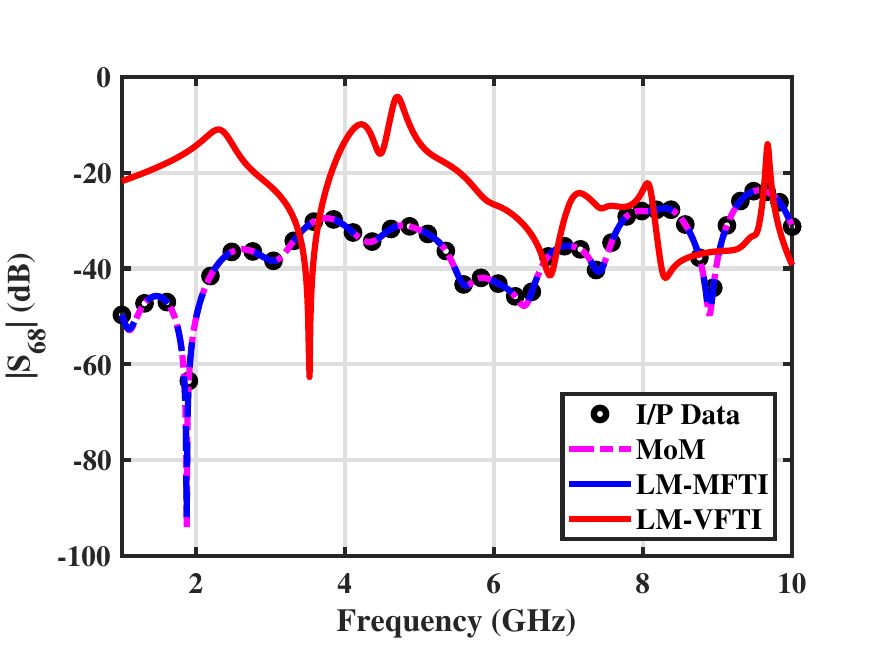}
      \caption{Comparison of the VFTI and MFTI for a ten-port data.}
   \label{fig:VFTI_MFTI}
\end{figure}
\subsubsection{Matrix Format Tangential Interpolation}
In MFTI, we utilize all the data points to predict the system. The right and left data matrices consist of right and left transfer function matrices given as 
\begin{align*}
\textbf{W} = [\textbf{H}_{a1}, ..., \textbf{H}_{an}]\in \mathbb{C}^{p\times np}
\end{align*}
\begin{align*}
\textbf{V}=\begin{bmatrix}
\textbf{H}_{b1}\\
\vdots\\
\textbf{H}_{bn}
\end{bmatrix}\in \mathbb{C}^{np\times p}
\end{align*}
We have implemented the MFTI by vectorization. Using the MFTI, we can write a state representation of the MIMO system as 
\begin{subequations}
\begin{align}
 \textbf{E}=-\mathbb{L}\in \mathbb{C}^{np\times np}\\
\textbf{A}=-\sigma\mathbb{L}\in \mathbb{C}^{np\times np}\\
 \textbf{B}=\textbf{V}\in \mathbb{C}^{np\times p}\\
 \textbf{C}=\textbf{W}\in \mathbb{C}^{p\times np}
\end{align}
\label{eq:loewner_state2}
\end{subequations}
where, the elements of $\mathbb{L}$ and $\sigma\mathbb{L}$ are block matrices, and  can be represented as
\begin{subequations}
    \begin{align}
    \mathbb{L}_{ij}=\frac{\textbf{H}_{aj}-\textbf{H}_{bi}}{s_{aj}-s_{bi}}\in \mathbb{C}^{p\times p}\\
    \sigma \mathbb{L}_{ij}=\frac{s_{aj}\textbf{H}_{aj}-s_{bi}\textbf{H}_{bi}}{s_{aj}-s_{bi}} \in \mathbb{C}^{p\times p}
\end{align}
\label{eq:loewner_MFTI}
\end{subequations}

A comparison of the two data selection schemes for a 10-port network is shown in Fig \ref{fig:VFTI_MFTI}, and it is observed that for a large number of ports, MFTI will be efficient since it takes all the matrix data. VFTI will lead to inaccurate results for high-dimensional data. 

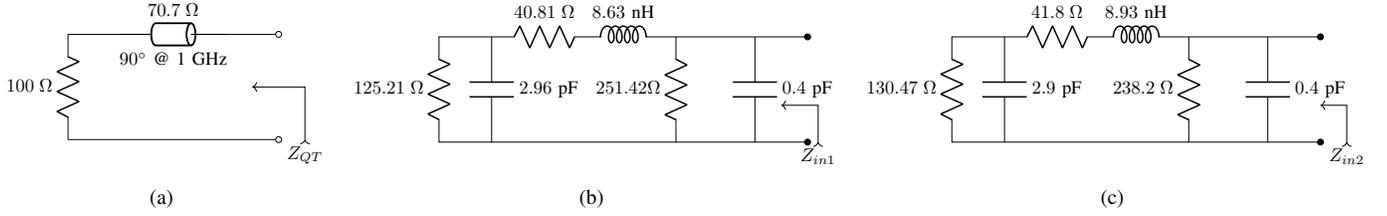
\begin{figure*}[!t]
\vspace{-1 em}
     \centering
     \subfloat[]{
     \begin{tikzpicture}[scale=0.9]
         \node[scale=0.70]{
         \begin{circuitikz}
    \draw (0,0) to[R=$100 \;\Omega$] (0,2)--(1,2);
    \draw (1,2) to[TL,l_={$90^{\circ}$ @ 1 GHz}] (3,2);
    \draw (1,2) to[TL,l={$70.7\;\Omega$}] (3,2);
    \draw (3,2) to[short,-o] (4,2);
    \draw (0,0) to[short,-o] (4,0); 
    \draw [->, to path={|- (\tikztotarget)}](4.5,0) edge (3.5,1);
    \draw (4.5,0) node[below]{$Z_{QT}$};
    \end{circuitikz}};
     \end{tikzpicture}\label{fig:CKT1}} 
     \subfloat[]{
     \begin{tikzpicture}[scale=0.9]
         \node[scale=0.70]{
\begin{circuitikz}
      \draw (0,0)
      to[R,l=$125.21 \;\Omega$] (0,2) % The voltage source
      to[short] (1,2)
      to[C=$2.96$ pF] (1,0) % The resistor
      to[short] (0,0);
       
      \draw (1,2)
 to [R=$40.81 \;\Omega$] (3,2) 
  to [L=$8.63$ nH] (4,2)  
  to[short] (4.5,2);
   \draw (4.5,2)
      to[R,l_=$251.42 \Omega$] (4.5,0) % The voltage source
      to[short] (1,0);
       \draw (4.5,2)
         to[short] (6,2);
      \draw (6,2)
       to[C,l=$0.4$ pF] (6,0) 
        to[short] (4.5,0);
         
           \draw (7,0) to [short, *-] (6,0);
        \draw (7,2) to [short, *-] (6,2);
 \draw [->, to path={|- (\tikztotarget)}](7.2,0) edge (6.5,0.7);
    \draw (7.2,0) node[below]{$Z_{in1}$};
        
\end{circuitikz}
         
         };
     \end{tikzpicture}\label{fig:CKT2}} 
     \subfloat[]{
     \begin{tikzpicture}[scale=0.9]
         \node[scale=0.70]{
\begin{circuitikz}
      \draw (0,0)
      to[R,l=$130.47 \;\Omega$] (0,2) % The voltage source
      to[short] (1,2)
      to[C=$2.9$ pF] (1,0) % The resistor
      to[short] (0,0);
       
      \draw (1,2)
 to [R=$41.8 \;\Omega$] (3,2) 
  to [L=$8.93$ nH] (4,2)  
  to[short] (4.5,2);
   \draw (4.5,2)
      to[R,l_=$238.2\; \Omega$] (4.5,0) % The voltage source
      to[short] (1,0);
       \draw (4.5,2)
         to[short] (6,2);
      \draw (6,2)
       to[C=$0.4$ pF] (6,0) 
        to[short] (4.5,0);
         
           \draw
      (7,0) to [short, *-] (6,0);
        \draw (7,2) to [short, *-] (6,2);
         \draw [->, to path={|- (\tikztotarget)}](7.5,0) edge (7,0.7);
    \draw (7.5,0) node[below]{$Z_{in2}$};
\end{circuitikz}         
         };
     \end{tikzpicture}\label{fig:CKT3}} 
     \caption{(a) Quarter Wave Transformer (QWT), (b) Equivalent Circuit-1, (c) Equivalent Circuit-2.}
    
 \end{figure*}
\begin{figure*}[!t]

\vspace*{-1em}
    \subfloat[]{\includegraphics[width=6cm]{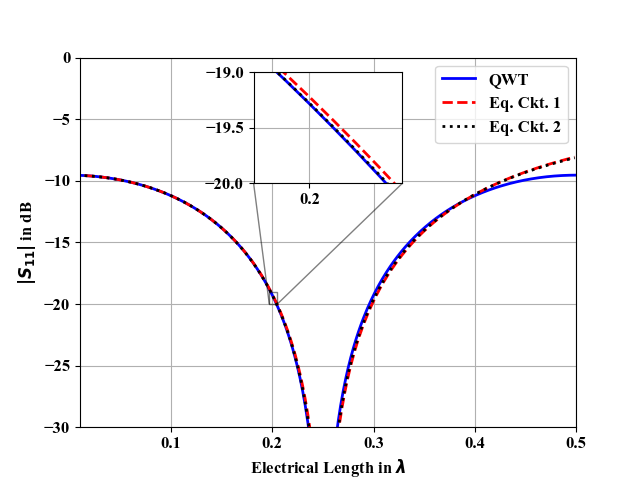}\label{fig:mag}}
    \subfloat[]{\includegraphics[width=6.25 cm]{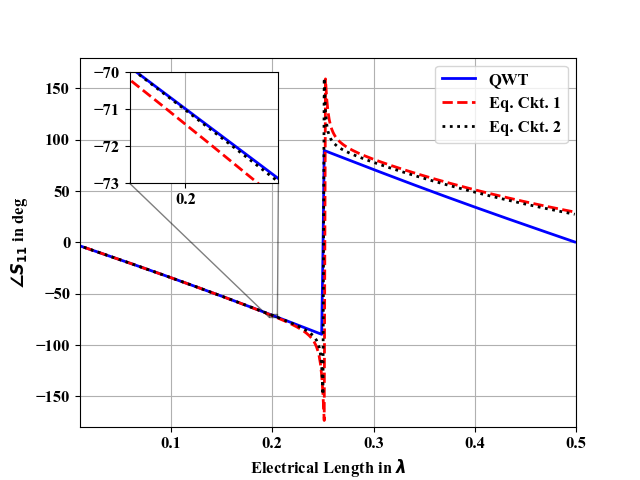}\label{fig:phase}}
    \subfloat[]{\includegraphics[width=6.25 cm]{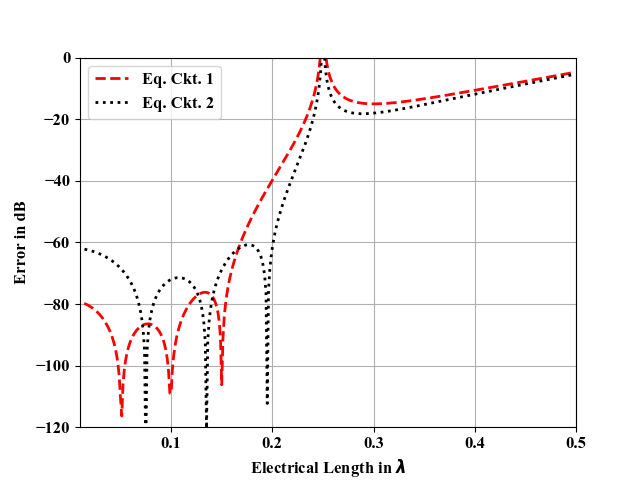}\label{fig:error}}
    \caption{Comparison of equivalent circuit models of QWT: (a) Magnitude (b) Phase (c) Error }
\end{figure*}

\section{Semi-adaptive and Adaptive Frequency Sweep using State Model\label{sec:salm}}
  
\subsection{Initial Order Selection from Electrical Size for Semi-adaptive Frequency Sweep \label{sec:order}}
 The vital step in the Loewner matrix framework for fast frequency sweep is finding the required frequency points, which will decide the initial order of the crude model. The initial order of the system was chosen by the most significant drop in the singular value plot of the matrix pencil \cite{Kabir_FFS1_2012} for the fast frequency sweep analysis. In the case of a distributed system, the drop will not be as significant as that of a lumped system, so the authors modified the algorithm and selected the determination of model order \cite{kabir2}, which is accomplished iteratively by considering the index of any of the five most substantial drops. The indices corresponding to the first five significant drops are arranged in descending order based on the drop values  \cite{kabir2}. However, this will only ensure an accurate model in limited cases.
  
 In recent work on fast frequency sweep \cite{Li_firstorder_2021}, the electrical size of 0.1$\lambda$ was approximated with a second-order circuit; therefore, to model a 0.2$\lambda$ structure, we need a fourth-order circuit. In this work, it has been shown that a 0.2$\lambda$ structure can be easily modelled as a third-order circuit. To demonstrate that, we have considered a quarter wave transformer which transforms $100\;\Omega$  load to a $50\; \Omega$  through a quarter wavelength transmission line of  0.25$\lambda$ at 1 GHz resonance frequency and having a characteristic impedance of 70.7 $\Omega$ shown in Fig. \ref{fig:CKT1}. Here, we have normalised the 1 GHz resonance frequency to 1 Hz. The input impedance  of the quarter-wave transformer can be expressed as
\begin{align}
     Z_{QT}(s)	= \frac {70.7 (170.7  e^{sk} + 29.3  e^{-sk})} {170.7  e^{sk} - 29.3  e^{-sk} } \label{eqQWT}.
\end{align}
where $ k=\frac {1}{4}$. We can understand that if we wish to exactly characterise the $Z_{QT}(s)$ from zero to infinite frequency, we need an infinite order circuit. If we expand $e^{sk}$ as a polynomial of $s$ using the Taylor series expansion, we need infinite order for the precision of $100\%$. However, for a finite frequency range and approximate accuracy, a finite-order system is good enough to characterize the system. Here, we will try to characterise the system as a third-order system.

% \begin{figure}[]
% \begin{center}
% \begin{circuitikz}
% \draw (0,0) to[R=$100 \Omega$] (0,2) to[short] (1,2);
% \draw (1,2) to[TL,l_={$90^{\circ}$ @ 1 GHz}] (3,2);
% \draw (1,2) to[TL,l={$70.7\;\Omega$}] (3,2);
% \draw (3,2) to[short] (4.5,2);
% \draw (0,0) to[short] (4.5,0); 
% \end{circuitikz}
% \end{center}
%\caption{Quarter Wave Transformer}
 %\label{fig:CTAT}
% \end{figure}
In order to verify the idea, we first fed the Loewner matrix framework with the input impedance of the quarter wave transformer at three discrete frequency points. The equivalent circuit model is created by converting the state-space model derived from the Loewner matrix to the system transfer function. Here, two examples of equivalent circuit models are given. In the first case, the input impedance of the transmission line is evaluated at frequencies 0.2 GHz, 0.4 GHz, and 0.6 GHz. The equivalent circuit of the system model obtained from the Loewner matrix is shown in Fig. \ref{fig:CKT2}, and the transfer function of the circuit can be expressed as
\begin{equation}
       Z_{in1}(s)	= \frac {2512 s^2 + 1.866\times 10^4 s + 1.305\times 10^5} {s^3 + 17.42 s^2 + 417.3 s + 1305 } \label{eqCKT2}.
         \end{equation}
In the second case, the input impedance of the transmission is evaluated at 0.3 GHz, 0.54 GHz, and 0.78 GHz frequency points. The equivalent circuit of the system model obtained from the Loewner matrix is shown in Fig. \ref{fig:CKT3}, and the transfer function of the circuit can be expressed as 
\begin{equation}
       Z_{in2}(s)	= \frac {2486 s^2 + 1.82\times 10^4 s + 1.267\times 10^5} {s^3 + 17.77 s^2 + 405.8 s + 1267} \label{eqCKT1}.
         \end{equation}
To validate the results, we have plotted the magnitude and phase of the S parameter for the quarter wave transformer and the two equivalent circuits obtained from the Loewner matrix framework as shown in Fig \ref{fig:mag} and Fig. \ref{fig:phase}. The error in the approximation is shown in Fig. \ref{fig:error}. It is evident from Fig. \ref{fig:error} that, with a proper selection of frequency points a $0.2\lambda$ distributed line can be modelled with a third-order equivalent circuit with an error tolerance of $-60$ dB. Please note that it is important to choose the locations of frequency samples wisely. To further reduce the approximation error, we need a more than third-order approximation.  

The number of internal states of the model depends upon the electrical size of the system at the highest frequency of interest \cite[Ch. 12.1 ]{passive}. We have chosen 0.2$\lambda$ as a third-order circuit in our work. We can imagine a longer TL as cascade connections of multiple 0.2$\lambda$ TL. According to the electrical size of the structure, we have approximated a third-order circuit at each 0.2$\lambda$ length, and the number of frequency points is calculated according to that. Let \textit{l} be the total physical length of PCB traces, then the initial number of frequency points required for the full-wave simulation ($n_0$) can be calculated as
\begin{equation}
n_0  =  \left \lceil{\frac{15lf_{max}}{pc}}\right \rceil.  \label{eq:n}
\end{equation}

where $c=3\times10^{8}$ m/s is the speed of light and $p$ is the number of ports. 
   
  The full-wave simulation is performed for the number of frequency points calculated as \eqref{eq:n}, and the data from the S parameter is given as input to the Loewner framework. Please note that for a $p$-port network, we can predict the maximum $n p$ order system for $n$ frequency samples.      

% \begin{figure}
%     \centering
%     \subfloat[]{\includegraphics[width= 8.5cm]{figures/FINAL_GRAPHS/cktcomp_mag.eps}} \label{fig:mag}\\
%     \subfloat[]{\includegraphics[width= 8.5 cm]{figures/FINAL_GRAPHS/cktcomp_phase.eps}} \label{fig:phase}\\
%     \caption{Comparison of magnitude and phase of the S-parameter of QWT with the two equivalent circuits (a) Magnitude Plot (b) Phase Plot}
%      \label{fig:eqv}
%    \end{figure}
\subsection{Initial Nonuniform Sampling for Semi-Adaptive Frequency Sweep \label{sec:sampling}}
    Uniform sampling is employed in \cite{Kabir_FFS1_2012} and \cite{kabir2}. For circuits with large electrical sizes requiring more data points at higher frequencies, we have employed logarithmic sampling, which takes more data samples at higher frequencies and is better than uniform sampling. Let $f_{max}$ and $f_{min}$ be the maximum and minimum frequency of simulation and $n_0$ be the number of frequency points calculated according to the third-order approximation. We can generate the frequency points as given in the equation below.
%     \begin{subequations}
%         \begin{align}
%             &f=f_{ref}-10^{ f_{min_{log}}: \frac{f_{max_{log}}-f_{min_{log}}}{n_0-1}:f_{max_{log}}}\label{log}\\
% \text{where}\nonumber\\
% &f_{ref}  = 2f_{max}+f_{min} \label{eqfref}\\
% &f_{min_{log}}=\log_{10}(2f_{max})\\
% &f_{max_{log}}=\log_{10}(f_{min}+f_{max})
% % f_{log} = f_{min_{log}}: \frac{f_{max_{log}}-f_{min_{log}}}{N+1}:f_{max_{log}}
%         \end{align}
%     \end{subequations}
    \begin{align}
        f_0=&2f_{max}+f_{min}\nonumber\\&-10^{\text{linspace}(\log_{10}(2f_{max}),\log_{10}(f_{max}+f_{min}),n_0)} \label{eq:f_0}
    \end{align}
Please note that the proposed logarithmic sampling is different from conventional logarithmic sampling  $f=10^{\text{linspace}(\log_{10} f_{min}, \log_{10} f_{max}, n)}$. In conventional logarithmic sampling, very few frequency samples are taken from the lower frequency of the TF, which leads to more approximation errors in the low-frequency part. The proposed logarithmic sampling makes a balanced choice between conventional logarithmic sampling and uniform sampling.        
    
 \begin{figure}[!h]
    
\centering
\includegraphics[width=7 cm]{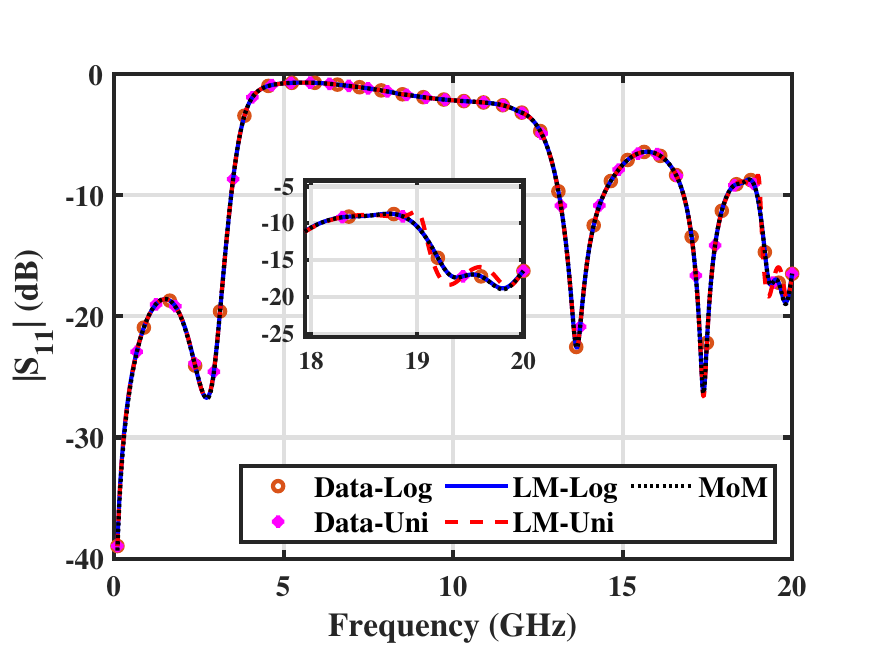}  
   \caption{Comparison of the two sampling scheme by plotting the magnitude  S$_{11}$ of a step impedance low pass filter. }
   \label{fig:com}
\end{figure}
To compare the two sampling techniques, the magnitude of the S$_{11}$ of a step impedance low pass filter is plotted in Fig. \ref{fig:com}. For comparison, we created the Loewner Matrix (LM) model for the two cases. In the first case, we have generated the $S_{11}$ data of the filter for uniformly sampled 36 frequency points. In the second case, the S-Parameter data of logarithmically sampled frequency points as in \eqref{eq:f_0} are generated. From Fig. \ref{fig:com}, it is clear that logarithmic sampling yields more accuracy than uniform sampling. Table \ref{table:2} compares the mean norm-2 error for both uniform and logarithmic sampling for various ports. The table makes it evident that logarithmic sampling has less error than uniform sampling. 
 \begin{table}[!h]
\begin{center}
 \caption{\label{demo-table} Comparison of the error for the two sampling schemes}
\label{table:2} 
\begin{tabular}{ |p{2cm}||p{2.5cm}||p{2.5cm}|} 
\hline
 \hline
   No of Ports & Error for Uniform Sampling &     Error for Logarithmic Sampling\\   
   \hline
   Two port   & 1.2 $\times$  10$^{-2}$ & 4.8 $\times$  10$^{-3}$ \\
 \hline
   Four port  & 1.6 $\times$  10$^{-3}$ & 4.11 $\times$ 10$^{-4}$\\
   \hline
  
 \end{tabular}
\end{center}
\end{table}

  \subsection{Data Partitioning Schemes for Loewner Matrix Formation \label{sec:partition}}
		After getting the frequency samples, the next step in the Loewner matrix method is to partition the frequency points and the corresponding data into left and right data sets. There are three types of data partitioning schemes, which are discussed in \cite{Chou_partition_2020}. It has been shown that the even-odd partitioning scheme can provide a better condition number of Lowener and shifted Lowener matrices. However, the even-odd partitioning scheme can not be applicable to an odd number of samples $n$. Therefore, we can use a combination of even-odd and positive-negative schemes for odd $n$ and even-odd schemes for even $n$. 

        If we arrange all the samples $f_0$ in ascending order $f_{01}<f_{02}<f_{03}<....<f_{0n}$, then we can arrange the data according to the proposed scheme as 
        \begin{small}
        \begin{subequations}
        \allowdisplaybreaks
        \begin{align}
          \textbf{s}_b=&\begin{cases} 
      [s_{1},\bar{s}_1,s_{3},\bar{s}_{3},....,s_{n-1},\bar{s}_{n-1}] & \text{if } n \text{ even} \\
      [s_1,\bar{s}_1,s_{3},\bar{s}_{3},....,s_{n-2},\bar{s}_{n-2},s_{n}] & \text{if } n \text{ odd} 
   \end{cases} \\
   \textbf{s}_a=&\begin{cases} 
      [s_2,\bar{s}_2,s_{4},\bar{s}_{4},....,s_{n},\bar{s}_{n}] & \text{if } n \text{ even} \\
      [s_1,\bar{s}_1,s_{3},\bar{s}_{3},....,s_{n-1},\bar{s}_{n-1},\bar{s}_{n}] & \text{if } n \text{ odd} 
   \end{cases} \\ 
   \textbf{H}_b=&\begin{cases} 
      [S_1,\bar{S}_1,S_{3},\bar{S}_{3},....,S_{n-1},\bar{S}_{n-1}] & \text{if } n \text{ even} \\
      [S_1,\bar{S}_1,S_{3},\bar{S}_{3},....,S_{n-2},\bar{S}_{n-2},S_{n}] & \text{if } n \text{ odd} 
   \end{cases} \\
   \textbf{H}_a=&\begin{cases} 
      [S_2,\bar{S}_2,S_{4},\bar{S}_{4},....,S_{n},\bar{S}_{n}] & \text{if } n \text{ even} \\
      [S_1,\bar{S}_1,S_{3},\bar{S}_{3},....,S_{n-1},\bar{S}_{n-1},\bar{S}_{n}] & \text{if } n \text{ odd} 
   \end{cases}  
        \end{align}   
        \label{eq:part}
        \end{subequations}
     \end{small}
where $s_k=j\omega_k=j2\pi f_{0k}$ and $S_k$ are $k^{th}$ complex frequency and scattering parameters of $k^{th}$ index. 

%		The achievement of accurate modelling in both the frequency and time domains is profoundly influenced by the choice of partitioning schemes, as highlighted in \cite{Chou_partition_2020}. Sub-optimal partitioning schemes impose severe limitations on the practical bandwidth achievable by the model. Furthermore, the condition number of the Loewner and shifted Loewner matrices substantially impacts the overall system performance. 
		To analyze the condition number, we have conducted assessments for both the $\mathbb{L}$ matrix and the $\sigma\mathbb{L}$ matrix for a ten port S-Parameter data, and the results are presented in Table \ref{table:1}. From the table, it is clear that in an even-odd partitioning scheme, the matrices exhibit good conditioning, as the denominator of the Loewner and shifted Loewner matrices remains relatively constant based on the partition. Consequently, this scheme minimizes significant variations in the data. Our experimental findings have led us to adopt the even-odd scheme as the preferred data partitioning approach for accurate predictions.
 \begin{table}[!h]
\begin{center}
\caption{\label{demo-table}Condition Number of the $\mathbb{L}$ and  $\sigma\mathbb{L}$ matrix in the context of Ten-port data.}
\label{table:1}
\begin{tabular}{ |p{2.5cm}||p{2.5cm}||p{2.5cm}|} 
\hline
 \hline
  Partition Scheme &    Condition number of $\mathbb{L}$ matrix &    Condition number of $\sigma\mathbb{L}$  matrix \\   
   \hline
   Even-Odd  & 5.61 $\times$ 10$^{13}$ & 3.19 $\times$ 10$^{13} $\\
 \hline

   Positive-Negative  & 3.96 $\times$ 10$^{19}$ & 1.17 $\times$ 10$^{19}$\\
   \hline
   
   High-Low & 1.79 $\times$ 10$^{19}$ & 1.29$\times$ 10$^{19}$ \\
   \hline 
 
  \end{tabular}

\end{center}
\end{table}

\subsection{ Model Order Reduction \label{sec:mor}}
	The reduced order model is generated using the singular value decomposition (SVD) approach \cite{S.Lefteriu_Loewner2010} . The singular value decomposition of the matrix  pencil $[x \mathbb{L}-\sigma\mathbb{L}] $ is done, expressed as 
 \begin{equation}
% [\textbf{Y},\Sigma,\textbf{X}] =	SVD (x\mathbb{L}-\sigma\mathbb{L})\label{eq}
x\mathbb{L}-\sigma\mathbb{L}=\textbf{Y}^{H}\mathbf{\Sigma}\textbf{X}\label{eq}
\end{equation}
  Given that $x$ can be any of the complex frequency points in the data set, and it should not be any of the eigenvalues of the matrix pencil $[\mathbb{L},\sigma\mathbb{L}] $. The reduced order ($r$) can be determined based on the singular value decay. By choosing the first $r$ columns of \textbf{X} and \textbf{Y} gives $\hat{\mathbf{X}}=\textbf{X}[:,1:r]$ and  $\hat{\mathbf{Y}}=\textbf{Y}[:,1:r]$ respectively, the reduced system realization can be represented as 
  \begin{subequations}
\allowdisplaybreaks
      \begin{align}
          \hat{\mathbf{E}}_r=&	-\hat{\mathbf{Y}}^{H}\mathbb{L}\hat{\mathbf{X}} \in\mathbb{C}^{r\times r}\label{eq}\\
          \hat{\mathbf{A}}_r =&	-\hat{\mathbf{Y}}^{H}\sigma\mathbb{L}\hat{\mathbf{X}}\in\mathbb{C}^{r\times r}\label{eq}\\
          \hat{\mathbf{B}}_r	= &\hat{\mathbf{Y}}^{H}\textbf{B}\in\mathbb{C}^{r\times p}\label{eq}\\
          \hat{\mathbf{C}}_r	=&\textbf{C} \hat{\mathbf{X}}\in\mathbb{C}^{p\times r}.\label{eq}
      \end{align}
      \label{eq:reduced_state}
  \end{subequations}
  Using the reduced order system, we can get the $r^{th}$ order approximation of the transfer function as 

  \begin{align}
    \hat{\textbf{H}}_r(s)=\hat{\textbf{C}}_r(s\hat{\textbf{E}}_r-\hat{\textbf{A}}_r)^{-1}\hat{\textbf{B}}_r.
    \label{eq:Hhat}
\end{align}
  The reduced model order selection is an essential step in the algorithm. The reduced order $r$ can be selected based on the following criterion 
  \begin{equation}
 \frac
                % Nominator
                {\sum\limits_{i=1}^r{\sigma}_{i}}
                % Denominator
                {\sum\limits_{i=1}^{np} {\sigma}_{i}} > 1-10^{-q}. \label{eq order}
\end{equation}
where $6\leq q \leq 12$ is a parameter that decides value of $r$ for large $n$. However, for small $n$, the parameter  $q$ does not have any effect.

\subsection{Adaptive selection of new samples \label{sec:adaptive}}
The problem with electrical length-based order selection is that (a) approximation error is decided based on $n_0$, and (b) the method fails for electrically large systems. In order to reduce the approximation error, we can add additional frequency points based on the error between two different order models with infinitesimally small frequency shifts. The next frequency point is decided based on the following pseudo error defined as 

\begin{align}
    E_{pseu.}(s)=|\hat{\textbf{H}}_{r_2}(s')-\hat{\textbf{H}}_{r_1}(s)|_2/ |\hat{\textbf{H}}_{r_1}(s)|_2 \label{eq:err_pseudo}
\end{align}
where $|\textbf{H}|_2=\text{max}(\sigma)$ with $\sigma$ is the singular value of matrix $\textbf{H}$. The variable $s'=j2\pi (f+\delta f)$ is shifted complex frequency. In this work, $\delta f=10^{-5}$ Hz has been considered for $f$ in Hz. The frequency perturbation helps to find new frequency samples when order reduction is not possible. The frequency perturbation relies on the continuity of TF. The reduced order $r_1$ and $r_2$ are selected based on two different $q=8$ and $q=12$ based on equation \eqref{eq order}. The new sample position can be obtained as 
\begin{align}
    f_{new}=\text{argmax}\left(E_{pseu.}(f)\right) \label{eq:fnew}
\end{align}
Simulate the system at a new frequency point $f_{new}$ and obtain actual TF $\textbf{H}(f_{new})$. We can get the actual error in the new frequency point as 
\begin{align}
    E_{act.}=|\hat{\textbf{H}}_{r_1}(s_{new})-\textbf{H}(s_{new})|_2/|\textbf{H}(s_{new})|_2\label{eq:err_act}
\end{align}
If the actual error $E_{act.}<tol$, we can say that the reduced order TF $\hat{\textbf{H}}_{r_1}(s)$ is converged to actual TF $\textbf{H}(s)$.  If the actual error $E_{act.}>tol$, append the new sample $\{s_{new}, \textbf{H}_{new}\}$ to an appropriate position in the existing data such that all the samples are arranged in ascending order. After appending the new sample to the existing data, we can apply the partition scheme \eqref{eq:part} to obtain the partitioned data $\{\textbf{s}_a, \textbf{H}_a\}$ and $\{\textbf{s}_b,\textbf{H}_b\}$. The partitioned data $\{\textbf{s}_a, \textbf{H}_a\}$ and $\{\textbf{s}_b,\textbf{H}_b\}$ are converted state matrices using \eqref{eq:loewner_state2} and \eqref{eq:loewner_MFTI}. The new state model $\{\textbf{E}, \textbf{A}, \textbf{B}, \textbf{C}\}$ is converted to two reduced state models $\{\hat{\textbf{E}}_{r_1}, \hat{\textbf{A}}_{r_1}, \hat{\textbf{B}}_{r_1},\hat{\textbf{C}}_{r_1}\}$  and $\{\hat{\textbf{E}}_{r_2}, \hat{\textbf{A}}_{r_2}, \hat{\textbf{B}}_{r_2},\hat{\textbf{C}}_{r_2}\}$ using \eqref{eq:reduced_state} and \eqref{eq order} for $q=8$ and $q=12$, respectively. Obtain two reduced-order TF model $\hat{\textbf{H}}_{r_1}(s)$ and $\hat{\textbf{H}}_{r_2}(s')$ for all testing frequency. Find the new frequency point $f_{new}$ using \eqref{eq:fnew} and obtain the actual error $E_{act.}$ using \eqref{eq:err_act}. The process will continue as long as $E_{act.}>tol$. If $E_{act.}<tol$ for three consecutive new samples, the process will be stopped and the final LM state model will be considered an approximation of the original TF.  The proposed technique is summarised as Algorithm-\ref{alg:cap}. Please note that we can use a frequency-independent desired $\textbf{D}$ matrix for better numerical stability.       In this work, the $\textbf{D}$ matrix has been considered as ones (i.e., $\textbf{D}=\text{ones}(p,p)$).      

\begin{algorithm}
\caption{Adaptive and Semi-adaptive frequency sweep using reduced-order LM state model}\label{alg:cap}
\begin{algorithmic}
\Require $f$, $\delta f$, $p$, $q_1$, $q_2$, $l$, $\textbf{Solver}$, D, sweep, $tol$ 
\Ensure $\textbf{H}(f)\approx \textbf{Solver}(f)$
\State $f_{min} \gets \text{min}(f)$
\State $f_{max} \gets \text{max}(f)$
\If{sweep is \textit{Adaptive}}
    \State $f_0 \gets [f_{min},f_{max}]$ 
\ElsIf{sweep is \textit{Semi-adaptive}}
    \State $n_0 \gets$ get from \eqref{eq:n}
    \State $f_0 \gets \text{get from }\eqref{eq:f_0}$ 
\EndIf
\State $H_0 \gets \textbf{Solver}(f_0)$
\State $H_0 \gets H_0-D$
\State $\{s_a, H_a\}, \{s_b, H_b\} \gets$ \text{get using} \eqref{eq:part}
\State $[E, A, B, C] \gets$ get using \eqref{eq:loewner_state2} and \eqref{eq:loewner_MFTI}
\State $E_{act} \gets 1$
\State $memory\gets 0$
\While{$ memory < 3$}
\State $r_1 \gets$ using \eqref{eq order} with $q_1$
\State $[E_{r1}, A_{r1}, B_{r1}, C_{r1}] \gets$ using \eqref{eq:reduced_state} with $r_1$
\State $H_{r1} \gets$ using \eqref{eq:Hhat} with $s=j2\pi f$
\State $r_2 \gets$ using \eqref{eq order} with $q_2$
\State $[E_{r2}, A_{r2}, B_{r2}, C_{r2}] \gets$ using \eqref{eq:reduced_state} with $r_2$
\State $H_{r2} \gets$ using \eqref{eq:Hhat} with $s=j2\pi (f+\delta f)$
\State $E_{pseu.} \gets$ using \eqref{eq:err_pseudo}
\State $f_{new}\gets \text{argmax}(E_{pseu.})$
\State $H_{new} \gets \textbf{Solver}(f_{new})$
\State $H_{new}\gets H_{new}-D$
\State $E_{act.}\gets$ using \eqref{eq:err_act} with $H_{new}$ and $H_{r1}(f_{new})$
\If{$E_{act}<=tol$}
\State $memory \gets memory +1$
\ElsIf{$E_{act}>tol$}
\State $memory \gets 0$
\EndIf
\State $f_0\gets \text{append}(f_0,f_{new})$
\State $H_0\gets \text{append}(H_0,H_{new})$
\State $\{s_a, H_a\}, \{s_b, H_b\} \gets$ \text{get using} \eqref{eq:part}
\State $[E, A, B, C] \gets$ get using \eqref{eq:loewner_state2} and \eqref{eq:loewner_MFTI}
\EndWhile
\State $r_1 \gets$ using \eqref{eq order} with $q_1$
\State $[E_{r1}, A_{r1}, B_{r1}, C_{r1}] \gets$ using \eqref{eq:reduced_state} with $r_1$
\State $\textbf{H}(f) \gets$ using \eqref{eq:Hhat} with $s=j2\pi f$
\State $\textbf{H}(f) \gets \textbf{H}(f)+D$
\end{algorithmic}
\end{algorithm}

We completed our discussion on the Loewner matrix-based fully-adaptive and semi-adaptive frequency sweep methods proposed in this work. In the following section, the effectiveness of the proposed methods in EM simulations has been discussed.

\section{RESULTS AND DISCUSSION}\label{sec:RESULTS}
\begin{figure*}[!t]
\vspace{-1em}
    \subfloat[]{\includegraphics[width= 8cm]{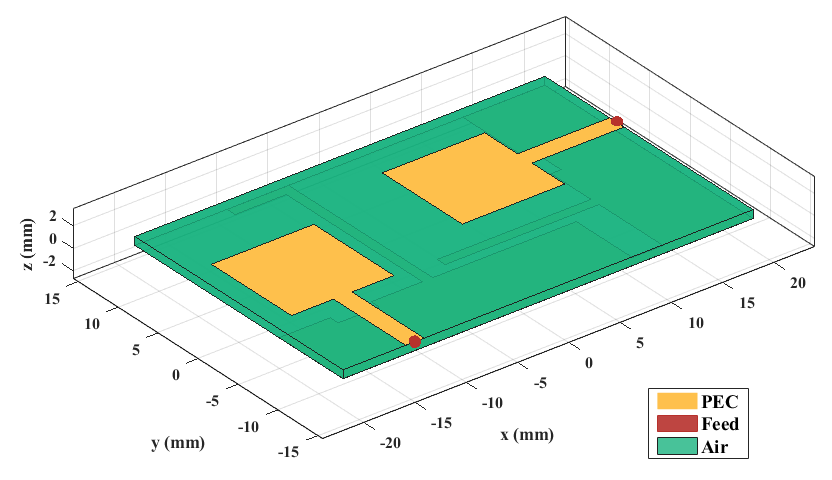} \label{fig:MIMO_antenna}} \hspace{4em}
    \subfloat[]{\includegraphics[width=8cm]{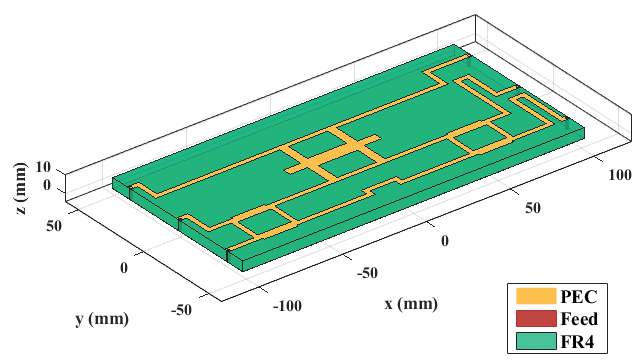}\label{fig:Nolen}}\\
    \subfloat[]{\includegraphics[width=8cm]{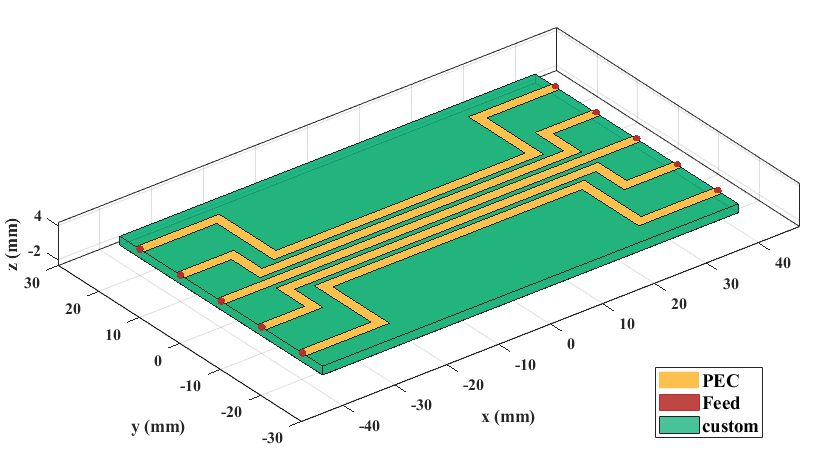}\label{fig:tenport}} \hspace{4em}
    \subfloat[]{\includegraphics[width=8cm]{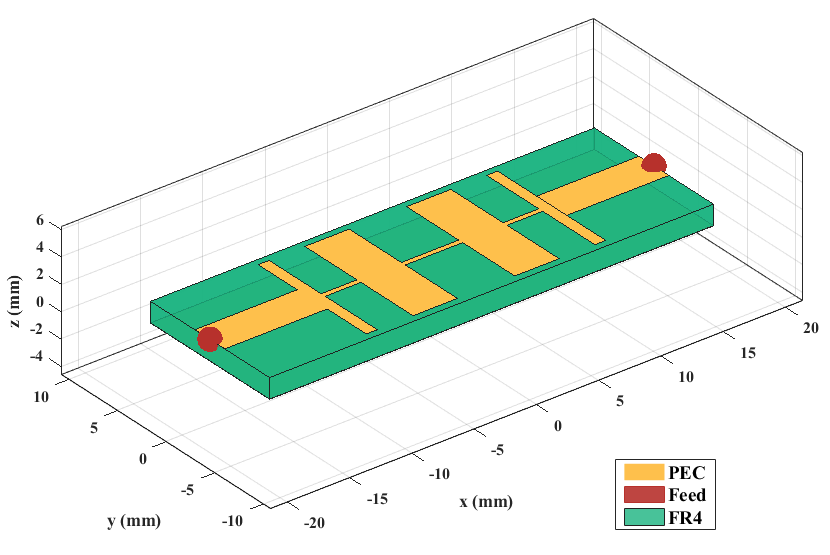}\label{fig:Step Impedance LPF}}
    \caption{(a) MIMO Antenna Array (b) Nolen Matrix (c) Ten Port PCB (d) Step Impedance LPF}
    \label{fig:pcb_exam}
\end{figure*}

In this section, four examples: (1) MIMO Antenna, (2) Nolen Matrix, (3) Ten-port Coupled-lines and (4) Step impedance Low Pass Filter are provided to illustrate the effectiveness of the fully adaptive and semi-adaptive Loewner matrix-based frequency sweep algorithm for electromagnetic simulation. \blue{Additional results for surface integrated waveguide (SIW), coplanar waveguide (CPW) and stripline are provided in the supplementary document\footnote[2]{A supplementary document of this paper is available online at http://ieeexplore.ieee.org \label{note1}}}.   The schematic diagrams of the example PCBs are shown in Fig. \ref{fig:pcb_exam}. The proposed algorithms have been compared with Pradovera's algorithm \cite{greedy} and the SB \cite{simo:24} algorithm. Brief descriptions of Pradovera's algorithm \cite{greedy} and the SB \cite{simo:24} algorithm have been discussed in Appendix-\ref{sec:prado} and Appendix-\ref{sec:SB}. Table \ref{tab:speci} shows different parameters like: no of ports $p$, frequency range, electrical size, full-wave simulation time and the total number of frequency sample $N$ for different examples. 
% The specifications of the examples are given in \ref{tab:speci}. 

For all four algorithms, we have considered $tol=-60$ dB and $memory =3$ (i.e., $E_{act}<tol$ for three consecutive new frequency points). The initial numbers of frequency points are 1, 2, 5, $n_0$ for Pradovera, fully adaptive LM, SB and semi-adaptive LM algorithms, respectively. The S-parameters data generated from the semi-adaptive LM algorithm are plotted for magnitude and phase response. The data generated from the model has been compared with the full-wave simulation of EM Data, which is obtained by simulating the structure with a 10 MHz interval over the bandwidth.  The error between the approximation models with 10 MHz interval and EM data have been compared in the error plot. The detailed description of the results is given in Tables \ref{tab:MIMO}, \ref{tab:Nolen}, \ref{tab:tenport} and \ref{tab:step}, respectively.
All the examples have been implemented using MATLAB RF PCB Toolbox with an HP z4 workstation. 
\begin{table*}[t]
    \centering
    \caption{Specification of the Examples}
    \label{tab:speci}
\begin{tabular}{ |p{4cm}||p{2cm}|p{2cm}|p{2cm}|p{2.5cm}|  }
\hline
\hline 
Example &  MIMO Antenna & Nolen Matrix & Ten Port PCB  & Step Impedance LPF\\
\hline
No of Ports $p$ &  2 & 6 & 10&  2\\ 
\hline
 Frequency Range ($f_{min}$-$f_{max}$) &  (1 - 8) GHz  & (1 - 4) GHz & (1 - 15) GHz& (1 - 30) GHz\\  
 \hline 
 Electrical Size ($\lambda$) &  1.06$\lambda$  & 2.71$\lambda$ & 4$\lambda$ & 3.57$\lambda$\\
  \hline 
 
   $t_{FW}$ (s) for 10MHz interval  & 1781.73 & 34897.85 & 48185.17 & 572.68\\\hline 
   Total samples $N$ &701 & 301&1401 &2901\\
    \hline
  %  Max Edge Length (m) &0.0208 & 0.0294& 0.0076 &0.0046\\
  % \hline
  % Minimum Edge Length (m) &0.0018 & 0.0036& 0.0020 &0.0031\\
  % \hline
  % Growth Rate &0.3400 & 0.5800& 0.6200 &0.7500\\
  % \hline
  
\end{tabular}
\end{table*}

\begin{figure*}[t]

\vspace*{-1.5em}
\centering
\subfloat[]{\includegraphics[width= 6cm]{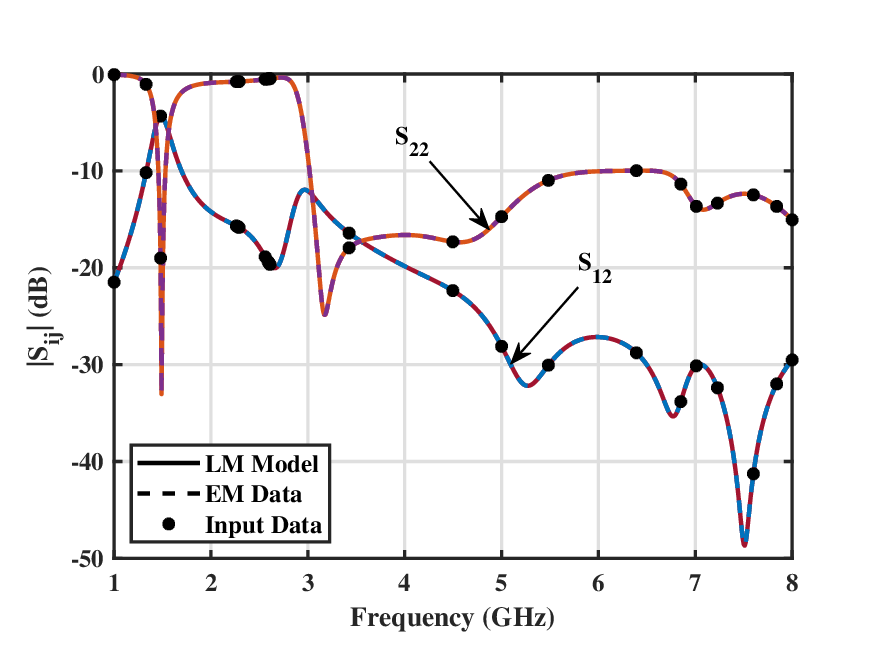}}
\subfloat[]{\includegraphics[width= 6cm]{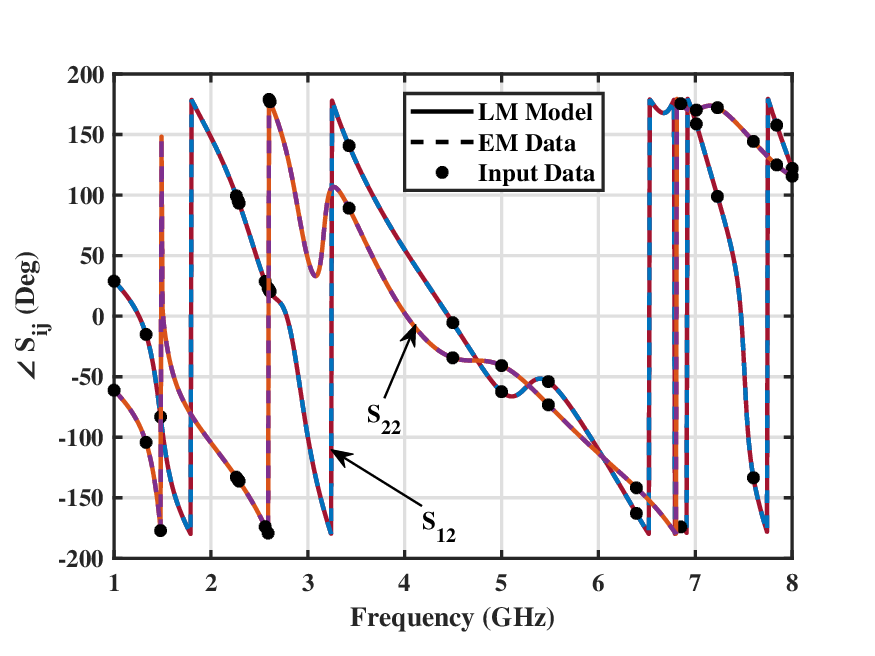}}
\subfloat[]{\includegraphics[width= 6 cm]{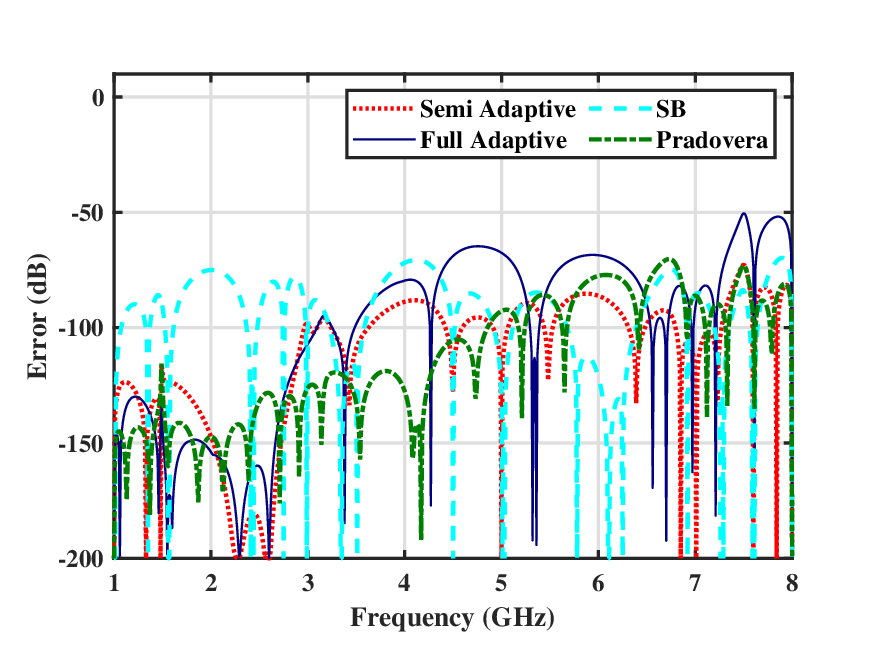}}
\caption{Simulation Results of MIMO Antenna (a) Magnitude Plot (b) Phase Plot (c) Error Plot}
\label{fig:mimoplot}
\end{figure*}

\begin{figure*}[t]

\vspace*{-1.5em}
    \centering
     \subfloat[]{\includegraphics[width= 6 cm]{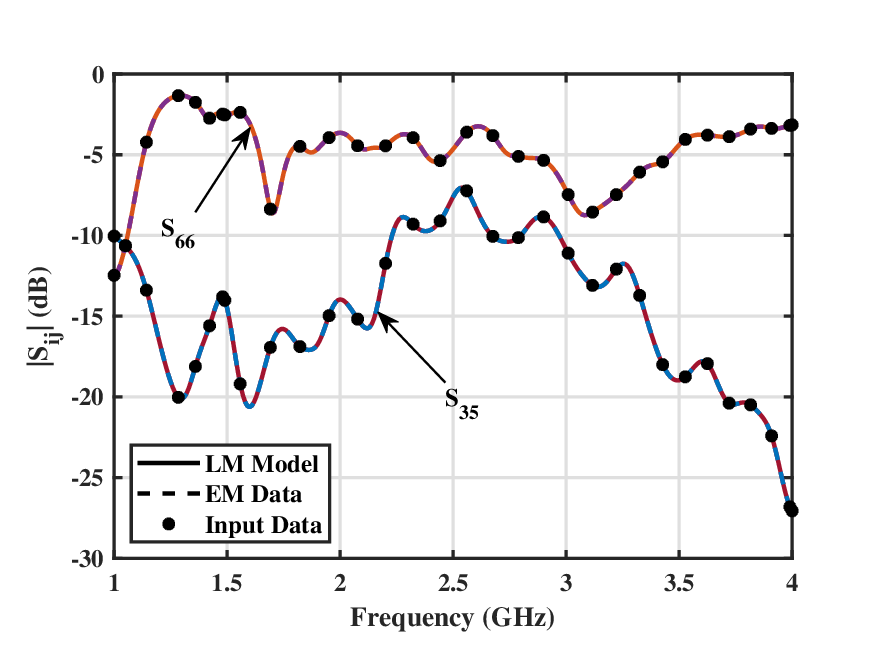}}
    \subfloat[]{\includegraphics[width= 6cm]{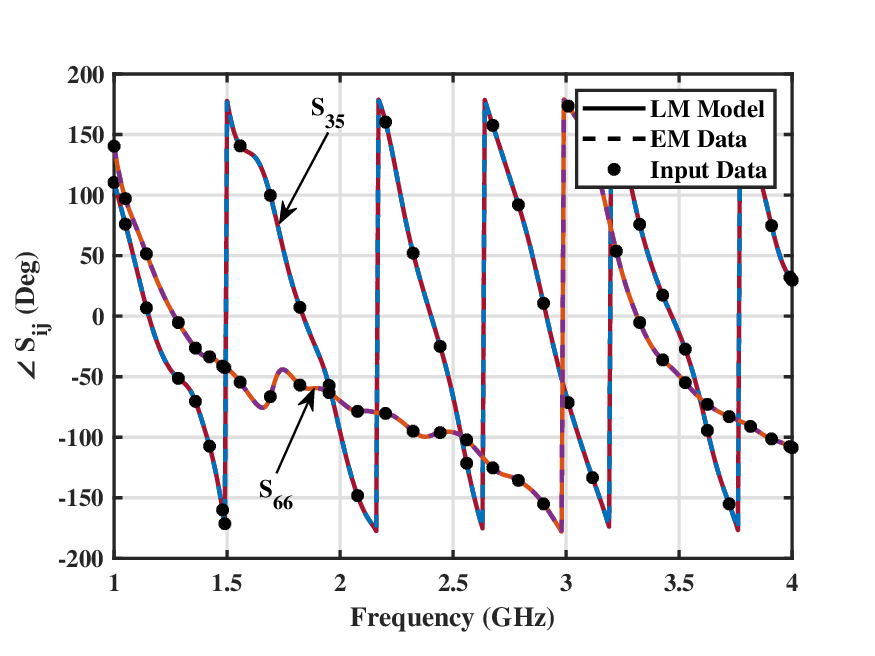}}
     \subfloat[]{\includegraphics[width= 6 cm]{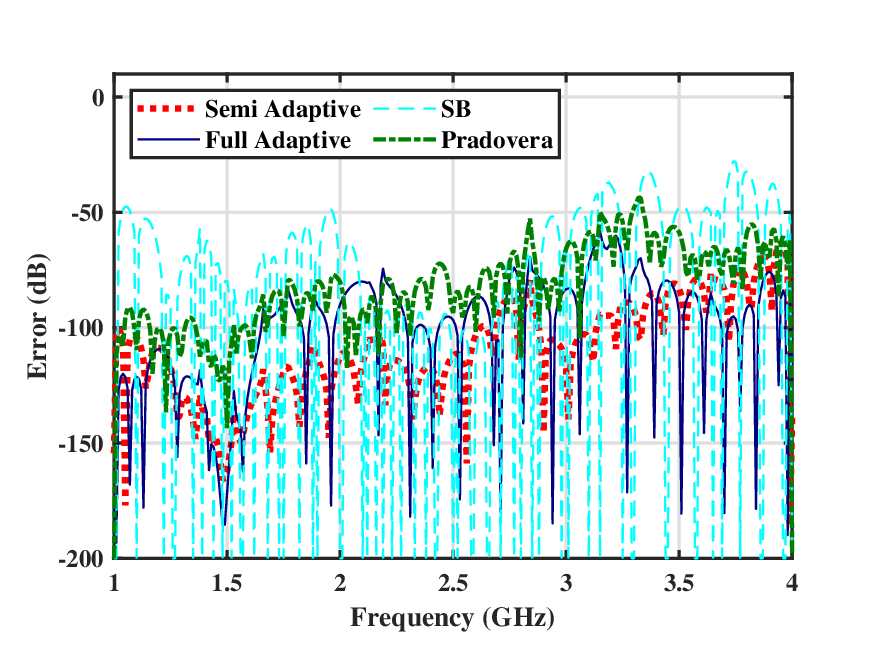}}
     \caption{Simulation Results of Nolen Matrix (a) Magnitude Plot (b) Phase Plot (c) Error Plot }
    \label{fig:nolenplot}
\end{figure*}

\begin{figure*}[t]

\vspace*{-1.5em}
    \centering
     \subfloat[]{\includegraphics[width= 6 cm]{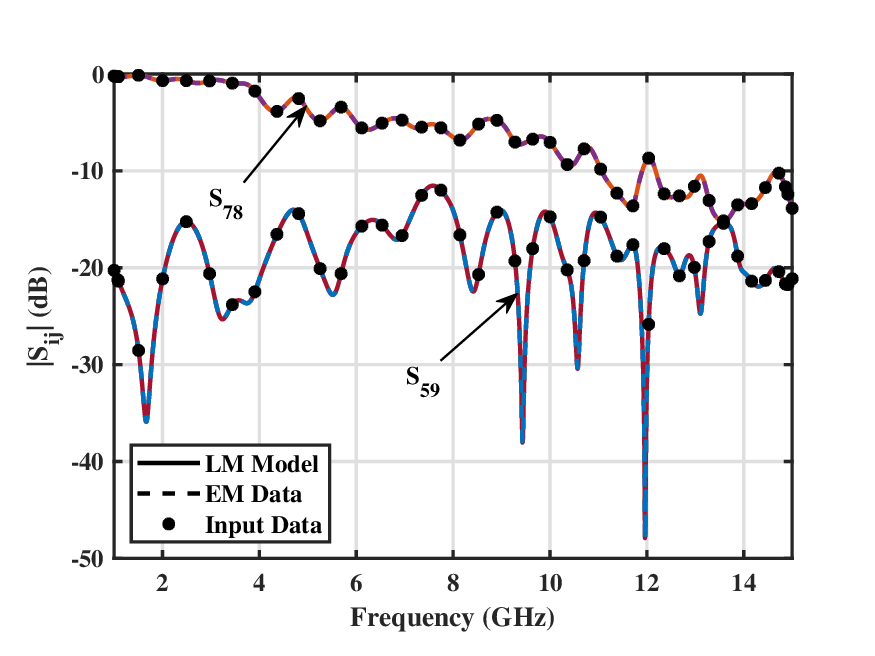}}
    \subfloat[]{\includegraphics[width= 6cm]{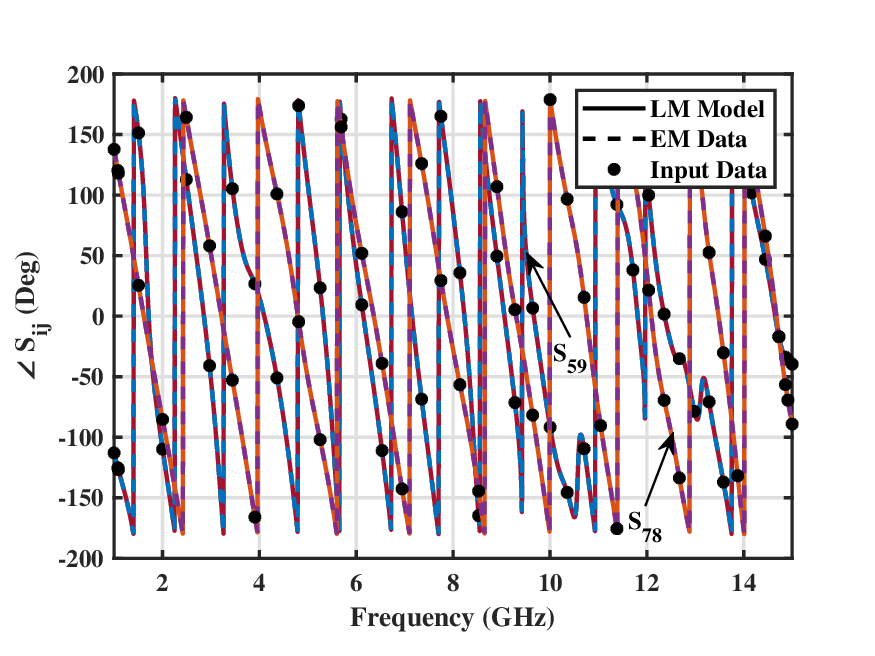}}
     \subfloat[]{\includegraphics[width= 6 cm]{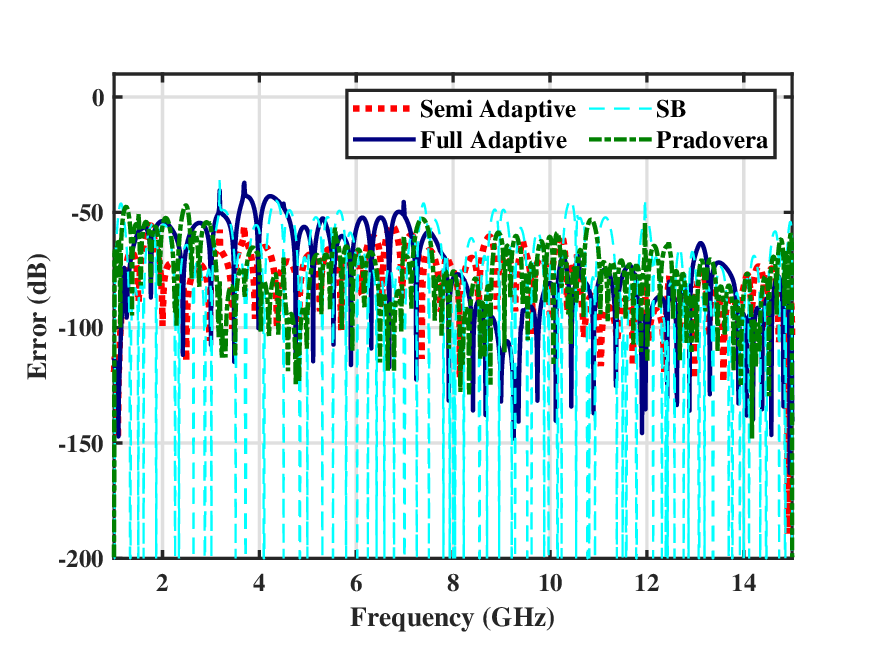}}
     \caption{Simulation Results of Ten port PCB (a) Magnitude Plot (b) Phase Plot (c) Error Plot }
    \label{fig:tenportplot}
\end{figure*}

\begin{figure*}[t]

\vspace{-1.5 em}
    \centering
     \subfloat[]{\includegraphics[width= 6 cm]{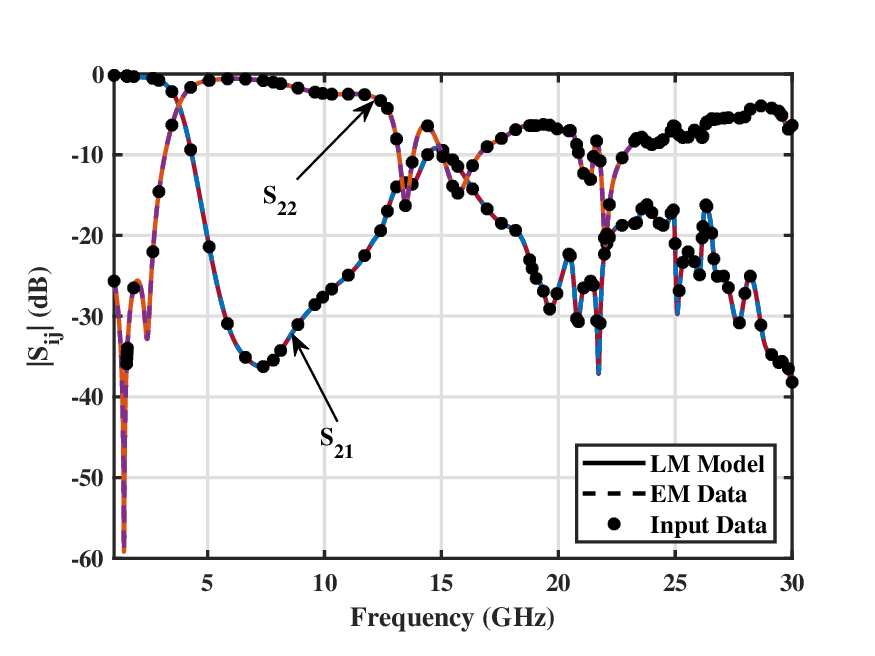}}
    \subfloat[]{\includegraphics[width= 6cm]{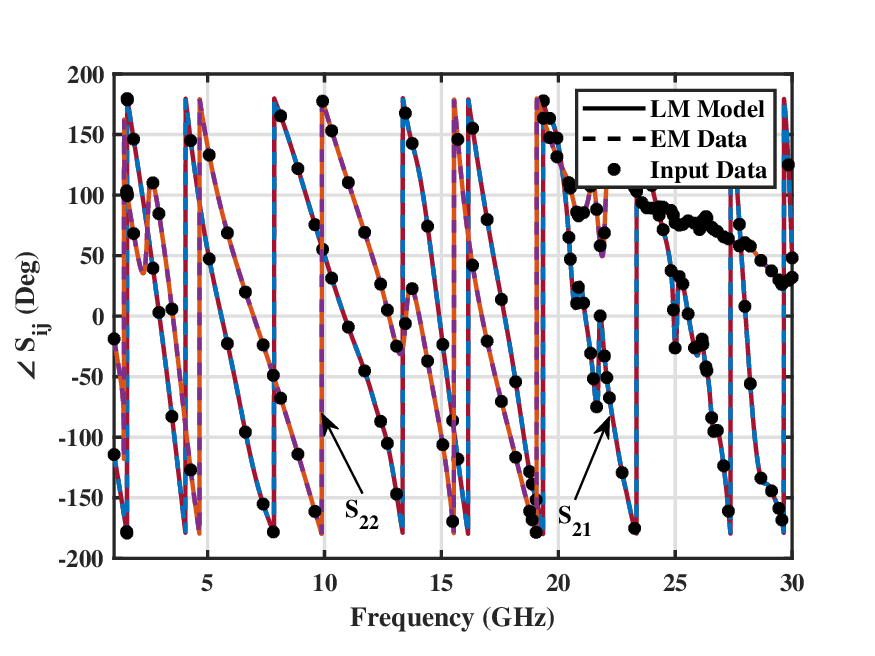}}
     \subfloat[]{\includegraphics[width= 6 cm]{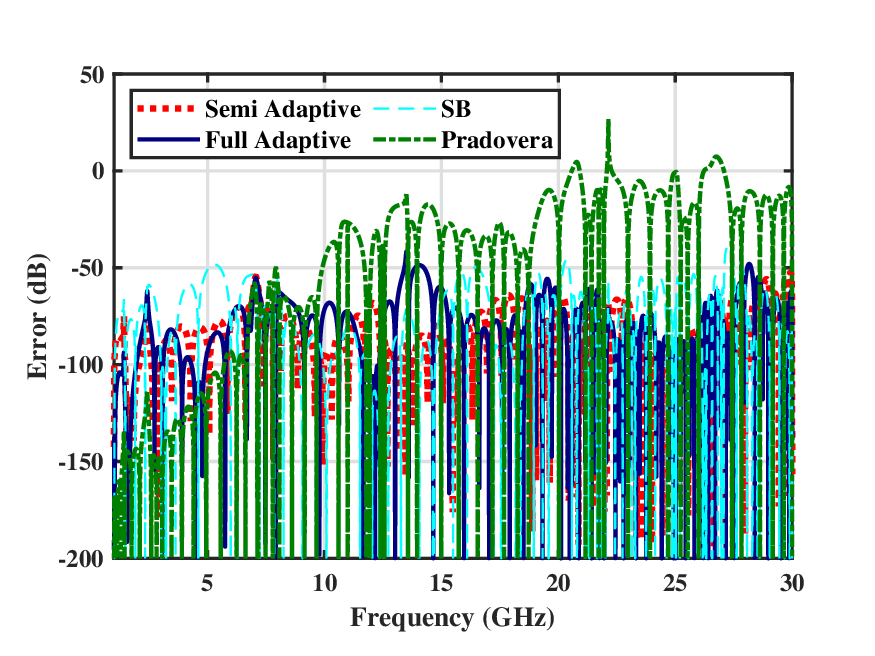}}
     \caption{Simulation Results of Step impedance low pass filter (a) Magnitude Plot (b) Phase Plot (c) Error Plot }
    \label{fig:twoportplot}
\end{figure*}

  \subsection{Example 1 - MIMO Antenna}
  In this example, a compact ultra-wideband (UWB) multiple-input multiple-output (MIMO) antenna array \cite{MIMO} has been simulated for adaptive frequency sweep. The MIMO array consists of two orthogonal planar monopole antennas lying in the horizontal plane shown in Fig. \ref{fig:MIMO_antenna}. The upper conductor of each monopole antenna measures 10 mm by 16 mm, while the conductor-backed dielectric substrate has dimensions of 40 mm by 26 mm. The ground layer is composed of two rectangular shapes, two rectangular slots, and three stubs. The MIMO array is designed on a dielectric substrate with a relative permittivity of 3.5 and a loss tangent of 0.004. A detailed description of the structure can be found in \cite{mimo_web}. The total length of the traces in the PCB is 40 mm. According to the proposed third-order approximation, $n_0=8$ frequency points are initially required for the semi-adaptive algorithm. The frequency range of simulation and electrical size of the structure, mesh data etc are given in table \ref{tab:speci}. The MIMO system has been simulated using the semi-adaptive, fully adaptive algorithm and also compared with SB and Pradovera's algorithm.  A detailed comparison of the results is tabulated in Table \ref{tab:MIMO}. From the table, we can see that the semi-adaptive LM method offers a better speed-up factor of 33.4 with less time and a maximum error of $-72.73$ dB for practical applications. The magnitude, phase and error plots of the MIMO antenna are shown in Fig. \ref{fig:mimoplot}. From the amplitude and phase response, it is evident that results obtained from AFS are indistinguishable from actual EM simulation results.

\begin{table}[t]
    \centering
    \caption{Adaptive Simulation Results of MIMO Antenna}
    \label{tab:MIMO}
\begin{tabular}{ |p{2.0cm}||p{1.3cm}|p{1.3cm}|p{0.8cm}|p{1.0cm}|  }
\hline
\hline 
Method &  Semi Adapt. LM & Fully Adapt. LM & SB \cite{simo:24} & Pradovera \cite{greedy}\\
\hline
Input samples $n$  &  21 & 21 & 25&  23\\ 
\hline
 $t_{adpt.}$ (s) &  53.25 & 62.41& 63.77& 60.66\\  
 \hline 
  max$(E_{act.})$ (dB) & \cellcolor{green!40}$-72.73$ &  \cellcolor{red!25}$-50.45$ & $-69$ & $-70.23$\\
  \hline 
 $SF=\frac{t_{FW}}{t_{adpt.}}$ & \cellcolor{green!40}33.4  & 28.5 & \cellcolor{red!25}27.9 & 29 \\
  \hline 
 \end{tabular}
\end{table}

\begin{table}[t]
    \centering
    \caption{Simulation Results of Nolen Matrix}
    \label{tab:Nolen}
\begin{tabular}{ |p{2.0cm}||p{1.3cm}|p{1.3cm}|p{0.8cm}|p{1.0cm}|  }
\hline
\hline 
Method &  Semi Adapt. LM & Fully Adapt. LM & SB \cite{simo:24}  & Pradovera \cite{greedy}\\
\hline
Input samples $n$  & 32 & 32 & 62 & 51  \\ 
\hline
 $t_{adpt.}$ (s) &  3316.60  & 3512.11& 6620.98& 5841.61\\  
 \hline 
  max$(E_{act.})$ (dB)&  \cellcolor{green!40}$-63.27$  & $ -58.42$  & \cellcolor{red!25}$-27.87$ & $-43.60$ \\
  \hline 
  $SF=\frac{t_{FW}}{t_{adpt.}}$ & \cellcolor{green!40}10.5  & 9.9 & \cellcolor{red!25}5.27 & 5.97 \\
  \hline 
 \end{tabular}
\end{table}

\begin{table}[t]
    \centering
    \caption{Simulation Results of Ten Port PCB}
    \label{tab:tenport}
\begin{tabular}{ |p{2.0cm}||p{1.3cm}|p{1.3cm}|p{0.8cm}|p{1.0cm}|  }
\hline
\hline 
Method &  Semi Adapt. LM & Fully Adapt. LM & SB \cite{simo:24}  & Pradovera \cite{greedy}\\
\hline
Input samples $n$  & 42 & 43 & 88 & 81  \\ 
\hline
 $t_{adpt.}$ (s) &  1447.62  & 1589.97 & 7845.20& 2643.51\\  
 \hline 
  max$(E_{act.})$ (dB) &  \cellcolor{green!40}$-55.39$  & \cellcolor{red!25}$-37.20$  & $-40$ & $-46.93$ \\
  \hline 
   $SF=\frac{t_{FW}}{t_{adpt.}}$ & \cellcolor{green!40}33.28  & 30.3 & \cellcolor{red!25}6 & 18.22 \\
  \hline
 \end{tabular}
\end{table}

\begin{table}[t]
    \centering
    \caption{Simulation Results of Step Impedance LPF}
    \label{tab:step}
\begin{tabular}{ |p{2.0cm}||p{1.3cm}|p{1.3cm}|p{0.8cm}|p{1.0cm}|  }
\hline
\hline 
Method &  Semi Adapt. LM & Fully Adapt. LM & SB \cite{simo:24}  & Pradovera \cite{greedy}\\
\hline
Input samples $n$  & 87 & 86 & 105 & 57 \\ 
\hline
  $t_{adpt.}$ (s) &  111.58  & 136.48& 1103.7& 12.20\\  
 \hline 
 max$(E_{act.})$ (dB) &  \cellcolor{green!40}$-52.04$ & $-41.72$  & $-36.02$ & \cellcolor{red!25}26.66 \\
  \hline 
  $SF=\frac{t_{FW}}{t_{adpt.}}$ & \cellcolor{green!40}5.13  & 4.19 & \cellcolor{red!25}0.5 & 46.94 \\
  \hline
 \end{tabular}
\end{table}

\subsection{Example 2 - Nolen matrix }
This example considers a 2D beam forming array using a planar Nolen matrix designed at 2 GHz with FR4 substrate \cite{nolen_web}. Three hybrid couplers and three phase shifters are arranged in a $3\times3$ Nolen matrix planar array with three input ports and three output ports. The substrate has a thickness of 1.6 mm and a loss tangent of 0.0026. The detailed description of this example is given in \cite{nolen_web}. The overall length of the trace in the PCB has been approximately calculated to be 820 mm. The frequency range of the simulation and electrical size of the structure are given in table \ref{tab:speci}. The semi-adaptive technique initially requires 27 frequency points, based on the third-order approximation. A detailed description of the results is shown in Table -\ref{tab:Nolen}. We may infer from the table that the semi-adaptive LM approach provides a high speedup factor with fewer input frequency points as well as better error performance. The magnitude, phase and error plots of the Nolen matrix are shown in Fig. \ref{fig:nolenplot}.

\subsection{Example 3 -Ten Port PCB }
A ten-port microstrip line PCB structure shown in Fig. \ref{fig:tenport} is simulated. The substrate is 1.58 mm thick with $\varepsilon_\mathrm{r} = 4.4 $, chosen as lossless. The length and width of the ground plane are 80 mm and 40 mm, respectively. The trace width is 2 mm, and there is a 2 mm gap between the traces. The simulation specifications are given in table \ref{tab:speci}. The total length of the trace is calculated as 500 mm. According to the third-order approximation 38 frequency points are required within the bandwidth for the semi-adaptive LM algorithm. Detailed comparisons of the results simulated using different algorithms are given in Table  \ref{tab:tenport}. Semi-adaptive LM provides a high speedup factor of 33.28 with a good maximum error of $-55.39$ dB. The fully adaptive LM approach also performs well in terms of the speedup factor and input data. The simulation results are plotted in Fig. \ref{fig:tenportplot}.

  \subsection{Example 4 - Step Impedance Low Pass Filter}
In this example, a seventh-order step impedance low pass filter is designed with a specified cut-off frequency of 5 GHz shown in  Fig. \ref{fig:Step Impedance LPF}. The port impedance is $50\;\Omega$, and the high and low impedances are taken as $120\;\Omega$ and $20\;\Omega$, respectively. The complete configuration is designed on an FR4 substrate with a dielectric constant of 4.4 and a thickness of 1.6 mm. The substrate is selected to incorporate loss with a loss-tangent value of 0.02. The dimensions of the ground plane are 35.7 mm in length and 12.2 mm in width. The total length of the trace is calculated as 65 mm. Table \ref{tab:speci} provides information on the simulation setup. According to the proposed third-order approximation, 48 frequency points are needed initially for the semi-adaptive frequency sweep algorithm. Table \ref{tab:step} compares four algorithms in terms of error, input samples and speedup factors. From the table, we can see that the semi-adaptive LM method and fully adaptive LM method perform better than the other methods. Even though the SB algorithm converges to the actual result, it is not suitable for AFS as the speed is less than the normal 10 MHz interval simulation time over the bandwidth. Therefore, the SB algorithm cannot be useful for the frequency sweep application. In this example, Pradovera's algorithm fails because the error is 26.66 dB. Therefore, we cannot ensure the convergence of Pradovera's algorithm for adaptive frequency sweep in all cases. The results are plotted in Fig.  \ref{fig:twoportplot}.

\subsection{Discussion}
This paper proposes a semi-adaptive and fully adaptive frequency sweep algorithm for electromagnetic simulation. The Loewner matrix framework has been exploited to generate two rational approximation models from the data. In the fully adaptive case, the maximum and minimum frequency of simulation is considered for the initial simulation. Further support points are calculated according to the proposed algorithm. In the semi-adaptive algorithm, the initial number of frequency points is calculated based
on the electrical size of the structure. The two algorithms have been compared with the existing SB algorithm and Pradovera's minimal sampling algorithm. Four examples are demonstrated with detailed simulation results shown in table \ref{tab:MIMO}, \ref{tab:Nolen}, \ref{tab:tenport} and table \ref{tab:step}. The results show that the proposed algorithms outperform the other two algorithms in terms of the speedup factor, error performance, and the requirement of minimal number of sample points. The fourth example demonstrates how Pradovera's method fails in terms of convergence and the SB approach fails in terms of speed. Overall, the results indicate that the proposed  Loewner matrix-based adaptive frequency sweep algorithm achieves faster simulation with fewer frequency points, maintaining good accuracy.

\section{CONCLUSION}\label{sec:Conclusion}
A Loewner matrix-based semi-adaptive frequency sweep algorithm has been proposed for EM simulation planar multiport PCBs. The initial number of samples is determined based on the electrical size of PCB traces with an approximation that there exists a $3^{rd}$ order equivalent circuit for $0.2\lambda$ line. To find additional frequency points to minimize actual error, two models with the same data are created using frequency perturbation and order reduction. The frequency points with the maximum pseudo error between the two models are chosen as the subsequent support points once the error between the models is computed. In this way, the algorithm chooses the critical frequency points adaptively within the bandwidth. The algorithm terminates when the actual error meets the specified error threshold with the help of memory. The proposed algorithm can also used in a fully adaptive sense, where two frequency points are chosen initially.  By comparing with the existing algorithm, the results show that the Loewner matrix-based adaptive frequency sweep approach performs better in terms of speed, accuracy and the requirement of minimum frequency points. With this approach, the speed of the electromagnetic simulation can be enhanced without compromising accuracy.

%\color{blue}
\section*{ACKNOWLEDGMENT}
The authors would like to sincerely thank the reviewers
and editors for their valuable suggestions to improve the
manuscript. The authors also thank D. Pradovera and P. S. Simon for making their works open source. The authors express a special thanks to Prof A. C. Antoulas and his associates for the development of the Loewner-State Model. 

\appendices

\renewcommand{\thefigure}{A\arabic{figure}}
\setcounter{figure}{0}
\renewcommand{\thetable}{A\arabic{table}} 
\setcounter{table}{0}
\renewcommand{\theequation}{A\arabic{equation}} 
\setcounter{equation}{0}

\section{Proof of Lowener-State Matrices \label{sec:proof}}
We can substitute $a$-set  $\{\textbf{s}_a, \textbf{H}_a\}$  and $b$-set $\{\textbf{s}_b, \textbf{H}_b\}$ data into \eqref{eq:Hs2}, which leads to 
\begin{subequations}
    \begin{align}
    \textbf{C}(s_{ai}\textbf{E}-\textbf{A})^{-1}\textbf{B}=H_{ai} \quad \forall i \in [1,n]\\
    \textbf{C}(s_{bj}\textbf{E}-\textbf{A})^{-1}\textbf{B}=H_{bj} \quad \forall j \in [1,n]
\end{align}
\label{eq:HaiHbj}
\end{subequations}

If we consider $\textbf{C}=\textbf{H}_a$ and $\textbf{B}=\textbf{H}_b$, then \eqref{eq:HaiHbj} can be written as 
\begin{subequations}
    \begin{align}
    \textbf{B}=(s_{ai}\textbf{E}-\textbf{A})\textbf{e}_i\label{eq:B}\\
   \textbf{C}=\textbf{e}^T_j(s_{bj}\textbf{E}-\textbf{A})\label{eq:C},
\end{align}
\label{eq:BC}
\end{subequations}
where $\textbf{e}_{i}=\mathds{1}[:,i]$ is the $i^{th}$ column of identity matrix of order $n$. After taking the inner product of \eqref{eq:B} and \eqref{eq:C} with respect to $\textbf{e}_{j}$ and $\textbf{e}_{i}$, leads to the following algebraic equations
\begin{subequations}
    \begin{align}
        H_{bj}=s_{ai}E_{ji}-A_{ji} \label{eq:EAa}\\
        H_{ai}=s_{bj}E_{ji}-A_{ji}\label{eq:EAb}
    \end{align}
\end{subequations}
where $E_{ji}=\textbf{E}[j,i]$ and $A_{ji}=\textbf{A}[j,i]$ are elements of $\textbf{E}$ and $\textbf{A}$ matrices. The linear equations \eqref{eq:EAa} and \eqref{eq:EAb} can easily be solved for $E_{ji}$ and $A_{ji}$, and solution is given as 
\begin{subequations}
    \begin{align}
        E_{ji}=-\frac{H_{bj}-H_{ai}}{s_{bj}-s_{ai}}=-\mathbb{L}_{ji}\\
        A_{ji}=-\frac{s_{bj}H_{bj}-s_{ai}H_{ai}}{s_{bj}-s_{ai}}=-\sigma\mathbb{L}_{ji}
    \end{align}
    \label{eq:Lij}
\end{subequations}
where $\mathbb{L}_{ji}$ and $\sigma\mathbb{L}_{ji}$ are elements of Loewner $\mathbb{L}$ and shifted Loewner $\sigma\mathbb{L}$ matrices. Therefore, one of the solutions of \eqref{eq:Hs2} is as in \eqref{eq:EABC}.
%\begin{align}
%    \textbf{E}=-\mathbb{L}; \hspace{1em}& \textbf{A}=-\sigma\mathbb{L}; \hspace{1em} &
%    \textbf{B}=\textbf{H}_b;\hspace{1em} & \textbf{C}=\textbf{H}_a. \label{eq:EABC}
%\end{align}
Therefore, the TF $H(s)$ can be represented in terms of sampled data as 
\begin{align}
   \widetilde{H}_n(s)=&\textbf{H}_a(-\mathbb{L}s+\sigma\mathbb{L})^{-1}\textbf{H}_b \nonumber\\
    =&\textbf{H}_a\textbf{W}_a\textbf{W}_a^{-1}(-\mathbb{L}s+\sigma\mathbb{L})^{-1}\textbf{W}_b^{-1}\textbf{W}_b\textbf{H}_b\nonumber\\
    =&\textbf{H}_a\textbf{W}_a(-\textbf{W}_b\mathbb{L}\textbf{W}_as+\textbf{W}_b\sigma\mathbb{L}\textbf{W}_a)^{-1}\textbf{W}_b\textbf{H}_b \label{eq:Hswawb}
\end{align}
After comparing \eqref{eq:Hs2} and \eqref{eq:Hswawb}, the general solution of \eqref{eq:Hs2} is given as in \eqref{eq:EABCg}.
%\begin{align}
%\begin{split}
%   \textbf{E}=-\textbf{W}_b\mathbb{L}\textbf{W}_a; \hspace{1em}& \textbf{A}=-\textbf{W}_b\sigma\mathbb{L}\textbf{W}_a;\\
%    \textbf{B}=\textbf{W}_b\textbf{H}_b;\hspace{1em} & \textbf{C}=\textbf{H}_a\textbf{W}_a. \label{eq:EABCg}
%\end{split}
%\end{align}

\section{ Pseudo-error based Adaptive Frequency Sweep for Barycentric approximation} \label{sec:prado}
In recent years, the barycentric approximation of transfer functions has gained popularity because of its stability and lower-order representation of transfer functions \cite{antoulas1986scalar},  \cite{AAA2018}. The $m^{th}$ order barycentric representation of a transfer function $H(s)$ can be written as 
\begin{align}
\widetilde{H}_m(s)=\frac{\sum_{j=1}^m\frac{w_jH(s_j)}{s-s_j}}{\sum_{j=1}^m\frac{w_j}{s-s_j}}
\end{align}
where where $s_1, . . . , s_m$ are a set of $m$ complex sample frequency with $s_j=j\omega_j=2\pi f_j$  and $w_1, . . . , w_m$ are a set of nonzero barycentric weights. Utilizing the conjugate symmetry property (i.e., $H(\bar{s})=\bar{H}(s)$), the weights ($w_j$) can be determined from the null space of Lowener Matrix formed using sample data points with positive-negative partitioning scheme. 

Let consider the approximation model $\widetilde{H}_m(s)$ is matched to the original TF $H(s)$ at $\bar{s}_i=-s_i$ frequency point with $H(\bar{s}_i)=\bar{H}(s_i)$. This matching leads to the following algebraic equation
\begin{align}
   \sum_{j=1}^m\frac{w_j\bar{H}(s_i)}{\bar{s}_i-s_j} -\sum_{j=1}^m\frac{w_jH(s_j)}{\bar{s}_i-s_j}=0.\label{eq:barymatch}
\end{align}
For $i=1,...,m$, \eqref{eq:barymatch} leads to following matrix equation
\begin{align}
     \mathbb{L}\textbf{w}=0
\end{align}
where $\textbf{w}=[w_1,....,w_m]^T$ is the column vectors of weights and $\mathbb{L}\in \mathbb{C}^{m\times m}$ is the Lowener matrix with 
\begin{align}
     \mathbb{L}_{ij}=\frac{\bar{H}(s_i)-H(s_j)}{\bar{s}_i-s_j}.
\end{align}
Therefore, $\textbf{w}$ is a vector corresponding to the null space of $\mathbb{L}$. If the Lowener matrix $\mathbb{L}$ is decomposed in singular value decomposition form as 
\begin{align}
    \mathbb{L}=\textbf{U}\mathbf{\Sigma}\textbf{V},
\end{align}
and singular values are arranged in descending order, then
\begin{align}
    \textbf{w}=\textbf{V}[:,m].
\end{align}
The sample frequency points are calculated adaptively to minimize the pseudo approximation error. The pseudo approximation error can be defined as \cite{greedy} 
\begin{align}
    \epsilon_m(s)=\frac{1}{\sum_{j=1}^m\frac{w_j}{s-s_j}}.
\end{align}
The $(m+1)^{th}$ frequency point $s_{m+1}$ or $f_{m+1}$ is selected as  $s_{m+1}=\text{argmax}(\epsilon_m(s)) $. 

The maximum order of approximation is selected from the tolerance of actual error. The order of the approximation will be selected as $n$,  if 
\begin{align}
    \frac{|\widetilde{H}_j(s_{j+1})-H(s_{j+1})|}{|H(s_{j+1})|}<tol
\end{align}
with $j=n$ for memoryless implementation. We can consider that the actual error is less than the tolerance for three consecutive frequency points ($j\in \{n-2, n-1, n\}$) for memory-based implementation. 

For MIMO system, the transfer functions $\textbf{H}(s_j)=\textbf{H}_j\in \mathbb{C}^{p\times p}$ for $j\in \{1,2,...,m\}$ are matrices. The matrices are arranged as vectors $\underbar{\textbf{H}}(s_j)=[\textbf{H}_j(:,1)^T,\textbf{H}_j(:,2)^T,...,\textbf{H}_j(:,p)^T]^T$ (i.e., $\underbar{\textbf{H}}(s_j)\in \mathbb{C}^{p^2\times 1}$).  Therefore, the size of the Loewner matrix for $m^{th}$ order approximation is $\mathbb{L}\in \mathbb{C}^{mp^2\times m}$. The block elements of the Loewner matrix are column vectors with
\begin{align}
     \mathbb{L}_{ij}=\frac{\bar{\underbar{\textbf{H}}}(s_i)-\underbar{\textbf{H}}(s_j)}{\bar{s}_i-s_j} \in \mathbb{C}^{p^2\times 1}.
\end{align}
The weight vector $\textbf{w}\in \mathbb{C}^{m\times1}$ can be determine from the null space of $\mathbb{L}\in \mathbb{C}^{mp^2\times m}$. 

\begin{figure}

\vspace*{-1em}
\centering
\includegraphics[width= 7 cm]{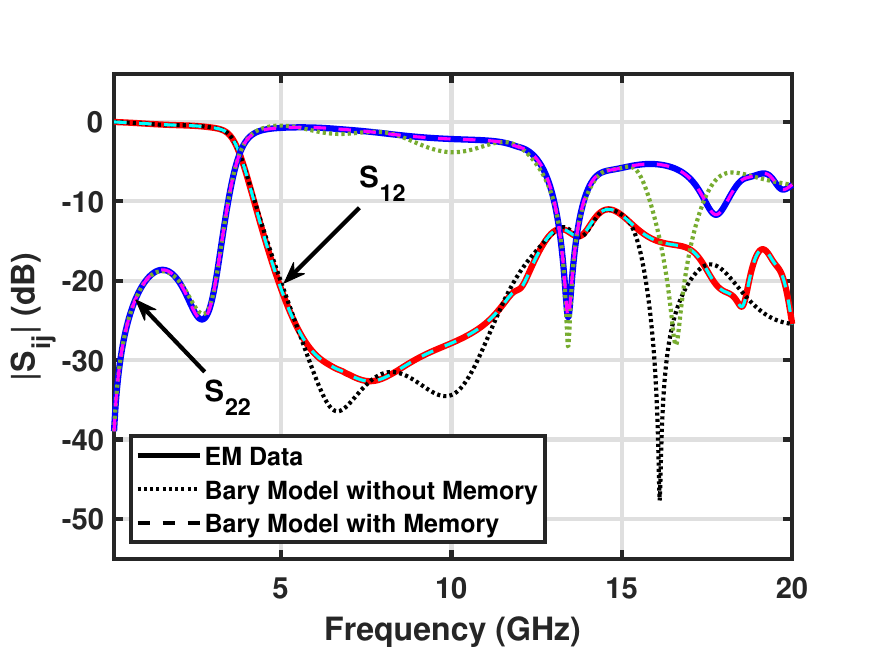}  \caption{Comparison of the Barycentric Model with and without memory }
   \label{fig:compare_mem}
\end{figure}

We will name this method as Pradovera's Method. Pradovera's Method was inspired by the AAA algorithm. In \cite{greedy}, Pradovera made two key contributions: (a) he introduced pseudo error in adaptive frequency find, and (b) he used the conjugate symmetry property of the transfer function in the formation of the Loewner matrix. Pradovera's Method has not been tested for the FFS of EM simulations. In this work, we tested Pradovera's Method for EM simulation. 
This algorithm can be applied both with and without memory. In some cases, the approximation converges without memory, but for certain structures, more frequency points are required, making the use of the algorithm with memory necessary. Fig.  \ref{fig:compare_mem}
 displays the S-parameter data of the step impedance low-pass filter demonstrated in Example 4 (refer Fig. \ref{fig:Step Impedance LPF}) using Pradovera's minimal sampling algorithm. Fig. \ref{fig:compare_mem} shows that the barycentric model without memory fails to converge to the full-wave simulation data. In this example, the algorithm without memory adaptively selects 20 frequency points, whereas 67 frequency points are needed when using memory. Although Pradovera's approach without memory uses fewer frequency points, it does not guarantee accuracy. Therefore, in this paper, we have opted for Pradovera's minimal sampling algorithm with memory for all examples and compared it to our proposed method.

\section{Stoer-Bulirsch Interpolating function based AFS \label{sec:SB}}
In recent work \cite{simo:24}, path-II of the Stoer-Bulirsch (SB) rational interpolation has been used for fast frequency sweep. The approximation of TF has been considered as a rational polynomial of $n^{th}$ order continued fraction $\widetilde{H}_n=H_{n1}$. The polynomial $H_{n1}$ can be obtained from a recursive relation as shown in Fig. \ref{fig:SB}. The elements of each column $\widetilde{H}_k$ with $k\in \{1,2,..., n\}$ in Fig. \ref{fig:SB} represent polynomials $H_{ki}$ of $k^{th}$ order. Polynomials $H_{ki}$ of the $k^{th}$ column are determined using the previous two columns. The recursive equation of the SB path-II algorithm can be written as 
\begin{align}
      H_{ki} =& H_{(k-2)(i+1)} + \frac{(f_{0(k+i)} - f_{0i}) d'_{ki} d''_{ki}}{n'_{ki} d''_{ki} + n''_{ki} d'_{ki}},
\end{align}
where
\begin{subequations}
    \begin{align}
        d'_{ki} =& H_{(k-1)(i-1)} - H_{(k-2)(i-1)}\\
        d''_{ki} =& H_{(k-1)i} - H_{(k-2)(i+1)}\\
        n'_{ki} =& f - f_{0i}\\
        n''_{ki} =& f_{0(k+i)} - f.        
    \end{align}
\end{subequations}
Here, $f_{0i}$ and $H_{0i}$ with $i\in \{1,2,..., n\}$ are the frequency and TF of the sample $i^{th}$. $H_{-1i}=0$ has been considered. Initially, the $H_{1i}$ polynomials are determined, and then subsequently the $H_{2i}$,..., $H_{ni}$ polynomials are determined.  
\begin{figure}
    \centering
\begin{tikzpicture}
    \node[scale=0.9]{$
\begin{array}{c|c|ccccc}
f_0 & H_0 & \widetilde{H}_1 & \widetilde{H}_2 & \cdots & \widetilde{H}_{n-1} &\widetilde{H}_{n} \\
\hline
f_{01} & H_{01} & H_{11} & H_{21} & \cdots & H_{(n-1)1} & H_{n1} \\
f_{02} & H_{02} & H_{12} & H_{22} & \cdots & H_{(n-1)2} \\
\vdots & \vdots & \vdots & \vdots & \ddots &  \\
f_{(n-2)} & H_{0(n-2)} & H_{1(n-2)} & H_{2(n-2)} & \cdots & \\
f_{(n-1)} & H_{0(n-1)} & H_{1(n-1)} & & & \\
f_n & H_{0n} & & & &
\end{array}
$};
\end{tikzpicture}
    \caption{Table for SB Algorithm}
    \label{fig:SB}
\end{figure}

The AFS starts with $n=5$ uniformly distributed frequency samples. New samples are added, where the pseudo error of the approximation is maximum. The pseudo error of this method is determined as 
\begin{align}
    E_{pseu.}=\frac{|H_{(n-1)1}-H_{n1}|}{|H_{n1}|}.
\end{align}
 The new frequency point is determined from pseudo error as $f_{new}=\text{argmax}( E_{pseu.})$. After getting TF at the new frequency point, the actual approximation error is determined as
 \begin{align}
     E_{act.}=\frac{|H(f_{new})-H_{n1}(f_{new})|}{|H(f_{new})|}.
 \end{align}
If the actual error $E_{act.}>tol$ add the sample to set $\{f_0, H_0\}$ and continue the AFS process. When $E_{act.}<tol$ for three consecutive new frequency samples the AFS will be terminated. Matrix or vector transfer functions can also be modelled using the SB algorithm by considering element-wise operations in the recursive formulation.

\color{black}
% \bibliographystyle{IEEEtran}
% \bibliography{REF}
\footnotesize
\printbibliography

@article{GURRALA2021107345,
title = {Comparison of vector and matrix format tangential interpolation for FDNE},
journal = {Electric Power Systems Research},
volume = {197},
pages = {107345},
year = {2021},
issn = {0378-7796},
doi = {https://doi.org/10.1016/j.epsr.2021.107345},
author = {Gurunath Gurrala and Kiran Kumar Challa}
}

@article{SALARIEH2021107254,
title = {Review and comparison of frequency-domain curve-fitting techniques: Vector fitting, frequency-partitioning fitting, matrix pencil method and loewner matrix},
journal = {Electric Power Systems Research},
volume = {196},
pages = {107254},
year = {2021},
issn = {0378-7796},
doi = {https://doi.org/10.1016/j.epsr.2021.107254},
author = {B. Salarieh and H.M.J. {De Silva}}
}

@ARTICLE{Gustavsen_VF99,
  author={Gustavsen, B. and Semlyen, A.},
  journal={IEEE Trans. Power Del.}, 
  title={Rational approximation of frequency domain responses by vector fitting}, 
  year={1999},
  volume={14},
  number={3},
  pages={1052-1061},
  doi={10.1109/61.772353}}

@ARTICLE{Convergency_VF2013,

  author={Lefteriu, Sanda and Antoulas, Athanasios C.},

  journal={IEEE Trans. Microw. Theory Tech.}, 

  title={On the Convergence of the Vector-Fitting Algorithm}, 

  year={2013},

  volume={61},

  number={4},

  pages={1435-1443},

  doi={10.1109/TMTT.2013.2246526}}

@ARTICLE{Convergency_VF2016,

  author={Shi, Guoyong},

  journal={IEEE Trans. Circuits Syst. II, Exp. Briefs}, 

  title={On the Nonconvergence of the Vector Fitting Algorithm}, 

  year={2016},

  volume={63},

  number={8},

  pages={718-722},

  doi={10.1109/TCSII.2016.2531127}}

@ARTICLE{S.Lefteriu_Loewner2010,
  author={Lefteriu, Sanda and Antoulas, Athanasios C.},
  journal={IEEE Trans. Computer-Aided Design Integr. Circuits Syst.}, 
  title={A New Approach to Modeling Multiport Systems From Frequency-Domain Data}, 
  year={2010},
  volume={29},
  number={1},
  pages={14-27},
  doi={10.1109/TCAD.2009.2034500}}

@ARTICLE{Burke_MBPE_1989,
  author={Burke, G.J. and Miller, E.K. and Chakrabarti, S. and Demarest, K.},
  journal={IEEE Trans. Magn.}, 
  title={Using model-based parameter estimation to increase the efficiency of computing electromagnetic transfer functions}, 
  year={1989},
  volume={25},
  number={4},
  pages={2807-2809},
  doi={10.1109/20.34291}}

@ARTICLE{Kottapalli1991,
  author={Kottapalli, K. and Sarkar, T.K. and Hua, Y. and Miller, E.K. and Burke, G.J.},
  journal={IEEE Trans. Microw. Theory Tech.}, 
  title={Accurate computation of wide-band response of electromagnetic systems utilizing narrow-band information}, 
  year={1991},
  volume={39},
  number={4},
  pages={682-687},
  doi={10.1109/22.76432}}

@INPROCEEDINGS{Wang_MFTI_2010,
  author={Wang, Yuanzhe and Lei, Chi-Un and Pang, Grantham K. H. and Wong, Ngai},
  booktitle={Design Automation Conference}, 
  title={MFTI: Matrix-format tangential interpolation for modeling multi-port systems}, 
  year={2010},
  volume={},
  number={},
  pages={683-686},
  doi={}}

@INPROCEEDINGS{Kabir_FFS1_2012,
  author={Kabir, Muhammad and Khazaka, Roni and Achar, Ramachandra and Nakhla, Michel},
  booktitle={2012 IEEE 21st Conference on Electrical Performance of Electronic Packaging and Systems}, 
  title={Loewner-Matrix based efficient algorithm for frequency sweep of high-speed modules}, 
  year={2012},
  volume={},
  number={},
  pages={185-188},
  doi={10.1109/EPEPS.2012.6457873}}

@INPROCEEDINGS{Chou_partition_2020,
  author={Chou, Chiu-Chih and Nguyen, Thong and Schutt-Ainé, José E.},
  booktitle={2020 IEEE Electrical Design of Advanced Packaging and Systems (EDAPS)}, 
  title={Impact of Partition Schemes in Loewner Matrix Macromodeling}, 
  year={2020},
  volume={},
  number={},
  pages={1-3},
  doi={10.1109/EDAPS50281.2020.9312918}}

@ARTICLE{Li_firstorder_2021,
  author={Li, Hongliang and Jin, Jian-Ming and Jachowski, Douglas R. and Hammond, Robert B.},
  journal={IEEE Trans. Microw. Theory Tech.}, 
  title={Fast Frequency Sweep Analysis of Passive Miniature RF Circuits Based on Analytic Extension of Eigenvalues}, 
  year={2021},
  volume={69},
  number={1},
  pages={4-14},
  doi={10.1109/TMTT.2020.3031584}}

@ARTICLE{Second-Order,
  author={Li, Hongliang and Jin, Jian-Ming and Jachowski, Douglas R. and Hammond, Robert B.},
  journal={IEEE Trans. Microw. Theory Tech.}, 
  title={Second-Order Analytic Extension of Eigenvalues for Fast Frequency Sweep Analysis of RF Circuits}, 
  year={2021},
  volume={69},
  number={4},
  pages={2078-2087},
  doi={10.1109/TMTT.2021.3056473}}

@ARTICLE{Zhu,
  author={Zhu, Kai and Wang, Jinhui and Yang, Shunchuan},
  journal={IEEE Antennas Wireless Propag. Lett.}, 
  title={An Adaptive Interpolation Scheme for Wideband Frequency Sweep in Electromagnetic Simulations}, 
  year={2022},
  volume={21},
  number={3},
  pages={471-475},
  doi={10.1109/LAWP.2021.3135958}}

@book{passive,
  title={Passive macromodeling: Theory and applications},
  author={Grivet-Talocia, Stefano and Gustavsen, Bjorn},
  year={2015},
  publisher={John Wiley \& Sons}}

@INPROCEEDINGS{kabir2,

  author={Kabir, Muhammad and Khazaka, Roni},

  booktitle={2014 IEEE 18th Workshop on Signal and Power Integrity (SPI)}, 

  title={Order selection for loewner matrix based macromodels for accurate macromodeling of distributed high-speed modules from limited number of full-wave S-parameter data}, 

  year={2014},

  volume={},

  number={},

  pages={1-4},

  doi={10.1109/SaPIW.2014.6844546}}

@ARTICLE{Matrix_pencil_TKS95,

  author={Sarkar, T.K. and Pereira, O.},

  journal={IEEE Antennas Propag. Mag.}, 

  title={Using the matrix pencil method to estimate the parameters of a sum of complex exponentials}, 

  year={1995},

  volume={37},

  number={1},

  pages={48-55},

  doi={10.1109/74.370583}
}

@ARTICLE{VF_Gustavsen99,

  author={Gustavsen, B. and Semlyen, A.},

  journal={IEEE Trans. Power Del.}, 

  title={Rational approximation of frequency domain responses by vector fitting}, 

  year={1999},

  volume={14},

  number={3},

  pages={1052-1061},

  doi={10.1109/61.772353}}

@ARTICLE{Kabir_Loewner_2012,
  author={Kabir, Muhammad and Khazaka, Roni},
  journal={IEEE Trans. Microw. Theory Tech.}, 
  title={Macromodeling of Distributed Networks From Frequency-Domain Data Using the Loewner Matrix Approach}, 
  year={2012},
  volume={60},
  number={12},
  pages={3927-3938},
  doi={10.1109/TMTT.2012.2222915}}

@ARTICLE{Keyhan_Matrix_Pencil2012,
  author={Sheshyekani, Keyhan and Karami, Hamid R. and Dehkhoda, Parisa and Paolone, Mario and Rachidi, Farhad},
  journal={IEEE Trans. Power Del.}, 
  title={Application of the Matrix Pencil Method to Rational Fitting of Frequency-Domain Responses}, 
  year={2012},
  volume={27},
  number={4},
  pages={2399-2408},
  doi={10.1109/TPWRD.2012.2208986}}

@Article{MartinezSpline2019,
AUTHOR = {Martinez, Juan A. and Belenguer, Angel and Esteban, Héctor},
TITLE = {Fast Frequency Sweep Technique Based on Segmentation for the Acceleration of the Electromagnetic Analysis of Microwave Devices},
JOURNAL = {Applied Sciences},
VOLUME = {9},
YEAR = {2019},
NUMBER = {6},
ARTICLE-NUMBER = {1118},
ISSN = {2076-3417},
DOI = {10.3390/app9061118}
}

@ARTICLE{Jeong_AWE2016,

  author={Jeong, Yi-Ru and Hong, Ic-Pyo and Lee, Kyung-Won and Lee, Jong-Hyun and Yook, Jong-Gwan},

  journal={IEEE Trans. Antennas Propag.}, 

  title={Fast Frequency Sweep Using Asymptotic Waveform Evaluation Technique and Thin Dielectric Sheet Approximation}, 

  year={2016},

  volume={64},

  number={5},

  pages={1800-1806},

  doi={10.1109/TAP.2016.2529681}}

@article{Dan_Jiao_AWE99,
author = {Jiao, Dan and Zhu, Xian-Yang and Jin, Jian-Ming},
title = {Fast and accurate frequency-sweep calculations using asymptotic waveform evaluation and the combined-field integral equation},
journal = {Radio Science},
volume = {34},
number = {5},
pages = {1055-1063},
doi = {https://doi.org/10.1029/1999RS900068},
year = {1999}
}

@ARTICLE{CFH1995,
  author={Chiprout, E. and Nakhla, M.S.},
  journal={IEEE Trans. Computer-Aided Design Integr. Circuits Syst.}, 
  title={Analysis of interconnect networks using complex frequency hopping (CFH)}, 
  year={1995},
  volume={14},
  number={2},
  pages={186-200},
  doi={10.1109/43.370425}}

@ARTICLE{FD_Prony2020,
  author={Ando, Shigeru},
  journal={IEEE Trans. Signal Process.}, 
  title={Frequency-Domain Prony Method for Autoregressive Model Identification and Sinusoidal Parameter Estimation}, 
  year={2020},
  volume={68},
  number={},
  pages={3461-3470},
  doi={10.1109/TSP.2020.2998929}}

@ARTICLE{Morales_FDF_2020,

  author={Morales Rodriguez, Jesus and Medina, Edgar and Mahseredjian, Jean and Ramirez, Abner and Sheshyekani, Keyhan and Kocar, Ilhan},

  journal={IEEE Trans. Power Del.}, 

  title={Frequency-Domain Fitting Techniques: A Review}, 

  year={2020},

  volume={35},

  number={3},

  pages={1102-1110},

  doi={10.1109/TPWRD.2019.2932395}}

@ARTICLE{Ruheli_Equivalent_ckt_1974,
  author={Ruehli, A.E.},
  journal={IEEE Trans. Microw. Theory Tech.}, 
  title={Equivalent Circuit Models for Three-Dimensional Multiconductor Systems}, 
  year={1974},
  volume={22},
  number={3},
  pages={216-221},
  doi={10.1109/TMTT.1974.1128204}}

@ARTICLE{AFS1991,

  author={Miller, E.K.},

  journal={Proceedings of the IEEE}, 

  title={Solving bigger problems-by decreasing the operation count and increasing the computation bandwidth}, 

  year={1991},

  volume={79},

  number={10},

  pages={1493-1504},

  doi={10.1109/5.104224}}

@Inbook{AFS1992,
author="Miller, E. K.
and Roberts, R. S.
and Charrabarti, S.",
editor="Brebbia, C. A.
and Ingber, M. S.",
title="Using Adaptive Frequency Sampling for More Efficient Determination of Broad Band Transfer Functions",
bookTitle="Boundary Element Technology VII",
year="1992",
publisher="Springer Netherlands",
address="Dordrecht",
pages="745--756",
isbn="978-94-011-2872-8",
doi="10.1007/978-94-011-2872-8_50",
}

@ARTICLE{AFSTKS1997,
  author={Adve, R.S. and Sarkar, T.K. and Rao, S.M. and Miller, E.K. and Pflug, D.R.},
  journal={IEEE Trans. Microw. Theory Tech.}, 
  title={Application of the Cauchy method for extrapolating/interpolating narrowband system responses}, 
  year={1997},
  volume={45},
  number={5},
  pages={837-845},
  doi={10.1109/22.575608}}

@ARTICLE{AFS1998,

  author={Peik, S.F. and Mansour, R.R. and Chow, Y.L.},

  journal={IEEE Trans. Microw. Theory Tech.}, 

  title={Multidimensional Cauchy method and adaptive sampling for an accurate microwave circuit modeling}, 

  year={1998},

  volume={46},

  number={12},

  pages={2364-2371},

  doi={10.1109/22.739224}}

@ARTICLE{AFS2001,
  author={Lehmensiek, R. and Meyer, P.},
  journal={IEEE Trans. Microw. Theory Tech.}, 
  title={Creating accurate multivariate rational interpolation models of microwave circuits by using efficient adaptive sampling to minimize the number of computational electromagnetic analyses}, 
  year={2001},
  volume={49},
  number={8},
  pages={1419-1430},
  doi={10.1109/22.939922}}

@ARTICLE{AFS2-2003,
  author={Yan Ding and Ke-Li Wu and Da Gang Fang},
  journal={IEEE Trans. Microw. Theory Tech.}, 
  title={A broad-band adaptive-frequency-sampling approach for microwave-circuit EM simulation exploiting Stoer-Bulirsch algorithm}, 
  year={2003},
  volume={51},
  number={3},
  pages={928-934},
  doi={10.1109/TMTT.2003.808694}}

@ARTICLE{AFS3-2008,
  author={Antonini, Giulio and Deschrijver, Dirk and Dhaene, Tom},
  journal={IEEE Trans. Electromagn. Compat.}, 
  title={Broadband Rational Macromodeling Based on the Adaptive Frequency Sampling Algorithm and the Partial Element Equivalent Circuit Method}, 
  year={2008},
  volume={50},
  number={1},
  pages={128-137},
  doi={10.1109/TEMC.2007.913225}}

@article{AFS4-2013,
author = {Li, Ping and Li, Yan and Jiang, Li Jun and Hu, Jun},
year = {2013},
month = {10},
pages = {5338-5343},
title = {A Wide-Band Equivalent Source Reconstruction Method Exploiting the Stoer-Bulirsch Algorithm With the Adaptive Frequency Sampling},
volume = {61},
journal = {IEEE Trans. Antennas Propag.},
doi = {10.1109/TAP.2013.2274032}
}

@ARTICLE{AFS-2024,
  author={Peumans, Dries and De Keersmaeker, Sander and Busschots, Cedric and Rolain, Yves and Ferranti, Francesco},
  journal={IEEE Trans. Microw. Theory Tech.}, 
  title={Adaptive Frequency Sampling Based on Local Rational Modeling for Microwave Electromagnetic Simulations}, 
  year={2024},
  volume={72},
  number={8},
  pages={4568-4578},
  doi={10.1109/TMTT.2024.3355270}}

@ARTICLE{LRM2021,

  author={Peumans, Dries and Pintelon, Rik and Lataire, John and Vandersteen, Gerd},

  journal={IEEE Trans. Instrum. Meas.}, 

  title={Frequency Response Function Measurements of Multivariable Systems via Local Rational Modeling}, 

  year={2021},

  volume={70},

  number={},

  pages={1-9},

  doi={10.1109/TIM.2021.3060590}}

@ARTICLE{AFS2019,
  author={Wu, Jun Wei and Cui, Tie Jun},
  journal={IEEE Access}, 
  title={Minimal Rational Interpolation and its Application in Fast Broadband Simulation}, 
  year={2019},
  volume={7},
  number={},
  pages={177813-177826},
  doi={10.1109/ACCESS.2019.2958369}}

@book{stoer1980introduction,
  title={Introduction to numerical analysis},
  author={Stoer, Josef and Bulirsch, Roland and Bartels, R and Gautschi, Walter and Witzgall, Christoph},
  volume={1993},
  year={1980},
  publisher={Springer}
}

@ARTICLE{MBPE1,

  author={Miller, E.K.},

  journal={IEEE Antennas Propag. Mag.}, 

  title={Model-based parameter estimation in electromagnetics. I. Background and theoretical development}, 

  year={1998},

  volume={40},

  number={1},

  pages={42-52},

  doi={10.1109/74.667326}}

@ARTICLE{MBPE2,
  author={Miller, E.K.},
  journal={IEEE Antennas Propag. Mag.}, 
  title={Model-based parameter estimation in electromagnetics. II. Applications to EM observables}, 
  year={1998},
  volume={40},
  number={2},
  pages={51-65},
  doi={10.1109/74.683542}}

@ARTICLE{MBPE3,
  author={Miller, E.K.},
  journal={IEEE Antennas Propag. Mag.}, 
  title={Model-based parameter estimation in electromagnetics. III. Applications to EM integral equations}, 
  year={1998},
  volume={40},
  number={3},
  pages={49-66},
  doi={10.1109/74.706084}}

@ARTICLE{AWE1990,

  author={Pillage, L.T. and Rohrer, R.A.},

  journal={IEEE Trans. Computer-Aided Design Integr. Circuits Syst.}, 

  title={Asymptotic waveform evaluation for timing analysis}, 

  year={1990},

  volume={9},

  number={4},

  pages={352-366},

  doi={10.1109/43.45867}}

@techreport{cockrell1996asymptotic,
  title={Asymptotic waveform evaluation (AWE) technique for frequency domain electromagnetic analysis},
  author={Cockrell, CR and Beck, FB},
  year={1996}
}

@ARTICLE{AWE1998,
  author={Reddy, C.J. and Deshpande, M.D. and Cockrell, C.R. and Beck, F.B.},
  journal={IEEE Trans. Antennas Propag.}, 
  title={Fast RCS computation over a frequency band using method of moments in conjunction with asymptotic waveform evaluation technique}, 
  year={1998},
  volume={46},
  number={8},
  pages={1229-1233},
  doi={10.1109/8.718579}}

@book{chiprout1994asymptotic,
  title={Asymptotic waveform evaluation},
  author={Chiprout, Eli and Nakhla, Michel S and Chiprout, Eli and Nakhla, Michel S},
  year={1994},
  publisher={Springer}
}

@article{Barry1986,
 ISSN = {00255718, 10886842},
 author = {Claus Schneider and Wilhelm Werner},
 journal = {Mathematics of Computation},
 number = {175},
 pages = {285--299},
 publisher = {American Mathematical Society},
 title = {Some New Aspects of Rational Interpolation},
 urldate = {2024-11-10},
 volume = {47},
 year = {1986}
}

@article{antoulas1986scalar,
  title={On the scalar rational interpolation problem},
  author={Antoulas, Athanasios C and Anderson, BDQ},
  journal={IMA Journal of Mathematical Control and Information},
  volume={3},
  number={2-3},
  pages={61--88},
  year={1986},
  publisher={Oxford University Press}
}

@article{AAA2018,
   title={The AAA Algorithm for Rational Approximation},
   volume={40},
   ISSN={1095-7197},
   DOI={10.1137/16m1106122},
   number={3},
   journal={SIAM Journal on Scientific Computing},
   publisher={Society for Industrial \& Applied Mathematics (SIAM)},
   author={Nakatsukasa, Yuji and Sète, Olivier and Trefethen, Lloyd N.},
   year={2018},
   month=jan, pages={A1494–A1522} }

@ARTICLE{StablePoleAAA2021,

  author={Valera-Rivera, Alvaro and Engin, Arif Ege},

  journal={IEEE Lett. Electromagn. Compat. Pract. Appl.}, 

  title={AAA Algorithm for Rational Transfer Function Approximation With Stable Poles}, 

  year={2021},

  volume={3},

  number={3},

  pages={92-95},

  doi={10.1109/LEMCPA.2021.3104455}}

@INPROCEEDINGS{Multi_function_AAA2020,
  author={Monzón, Lucas and Johns, William and Iyengar, Spatika and Reynolds, Matthew and Maack, Jonathan and Prabakar, Kumaraguru},
  booktitle={2020 IEEE Power \& Energy Society General Meeting (PESGM)}, 
  title={A Multi-function AAA Algorithm Applied to Frequency Dependent Line Modeling}, 
  year={2020},
  volume={},
  number={},
  pages={1-5},
  doi={10.1109/PESGM41954.2020.9281536}}

@INPROCEEDINGS{FastAAA2017,

  author={Hochman, Amit},

  booktitle={2017 IEEE 26th Conference on Electrical Performance of Electronic Packaging and Systems (EPEPS)}, 

  title={FastAAA: A fast rational-function fitter}, 

  year={2017},

  volume={},

  number={},

  pages={1-3},

  doi={10.1109/EPEPS.2017.8329756}}

@article{MAYO2007634,
title = {A framework for the solution of the generalized realization problem},
journal = {Linear Algebra and its Applications},
volume = {425},
number = {2},
pages = {634-662},
year = {2007},
note = {Special Issue in honor of Paul Fuhrmann},
issn = {0024-3795},
doi = {https://doi.org/10.1016/j.laa.2007.03.008},
author = {A.J. Mayo and A.C. Antoulas}
}

@article{greedy,
   title={Toward a certified greedy Loewner framework with minimal sampling},
   volume={49},
   ISSN={1572-9044},
   DOI={10.1007/s10444-023-10091-7},
   number={6},
   journal={Advances in Computational Mathematics},
   publisher={Springer Science and Business Media LLC},
   author={Pradovera, Davide},
   year={2023},
   month=dec }

@INPROCEEDINGS{MIMO,
  author={Liu, L. and Cheung, S. W. and Yuk, T. I. and Wu, D.},
  booktitle={2013 7th European Conference on Antennas and Propagation (EuCAP)}, 
  title={A compact ultrawideband MIMO antenna}, 
  year={2013},
  volume={},
  number={},
  pages={2108-2111},
  doi={}}

@online{mimo_web,
  author    = {Mathworks},
  title     = {Design and Analysis of Compact Ultra-Wideband MIMO Antenna Array},
 
  url       = {https://in.mathworks.com/help/antenna/ug/directivity-gain-and-realized-gain-of-ultrawideband-mimo-array.html},
  
}

@online{nolen_web,
  author    = {Mathworks},
  title     = {Analyze Nolen Matrix for the 2-D Beamforming Application},
 
  url       = {https://in.mathworks.com/help/rfpcb/ug/_mw_bff837db-5e26-4a56-b101-dd1a2ab3e1a1.html},
  
}

@article{simo:24,
  title={{PSSFSS}---An Open-source Code for Analysis of Polarization and Frequency Selective Surfaces},
  volume={39},
  DOI={https://doi.org/10.13052/2024.ACES.J.390207},
  number={02},
  journal={The Applied Computational Electromagnetics Society Journal (ACES)},
  author={Simon, Peter S.},
  year={2024},
  month={feb},
  pages={139--148}
}

@misc{Lucas24,
      title={On Adaptive Frequency Sampling for Data-driven MOR Applied to Antenna Responses}, 
      author={Lucas Åkerstedt and Darwin Blanco and B. L. G. Jonsson},
      year={2024},
      eprint={2409.18734},
      archivePrefix={arXiv},
      primaryClass={eess.SY}, 
}

@online{siw,
  author    = {Mathworks},
  title     = {Prototype, Design, and Analysis of SIW based Microstrip Tapered Transmission Line},
 
  url       = {https://in.mathworks.com/help/rfpcb/ug/prototype-design-and-analysis-of-siw-based-microstrip-tapered-transmission-line.html},
  
}

@online{CPW,
  author    = {Mathworks},
  title     = {Coplanar Waveguide},
 
  url       = {https://in.mathworks.com/help/rfpcb/ref/coplanarwaveguide.html},
  
}

@online{stripline,
  author    = {Mathworks},
  title     = {stripline},
 
  url       = {https://in.mathworks.com/help/rfpcb/ref/stripline.html},
  
}
\AtNextBibliography{\footnotesize}
\begin{IEEEbiography}[{\includegraphics[width=1 in,height=1.1 in, clip,keepaspectratio]{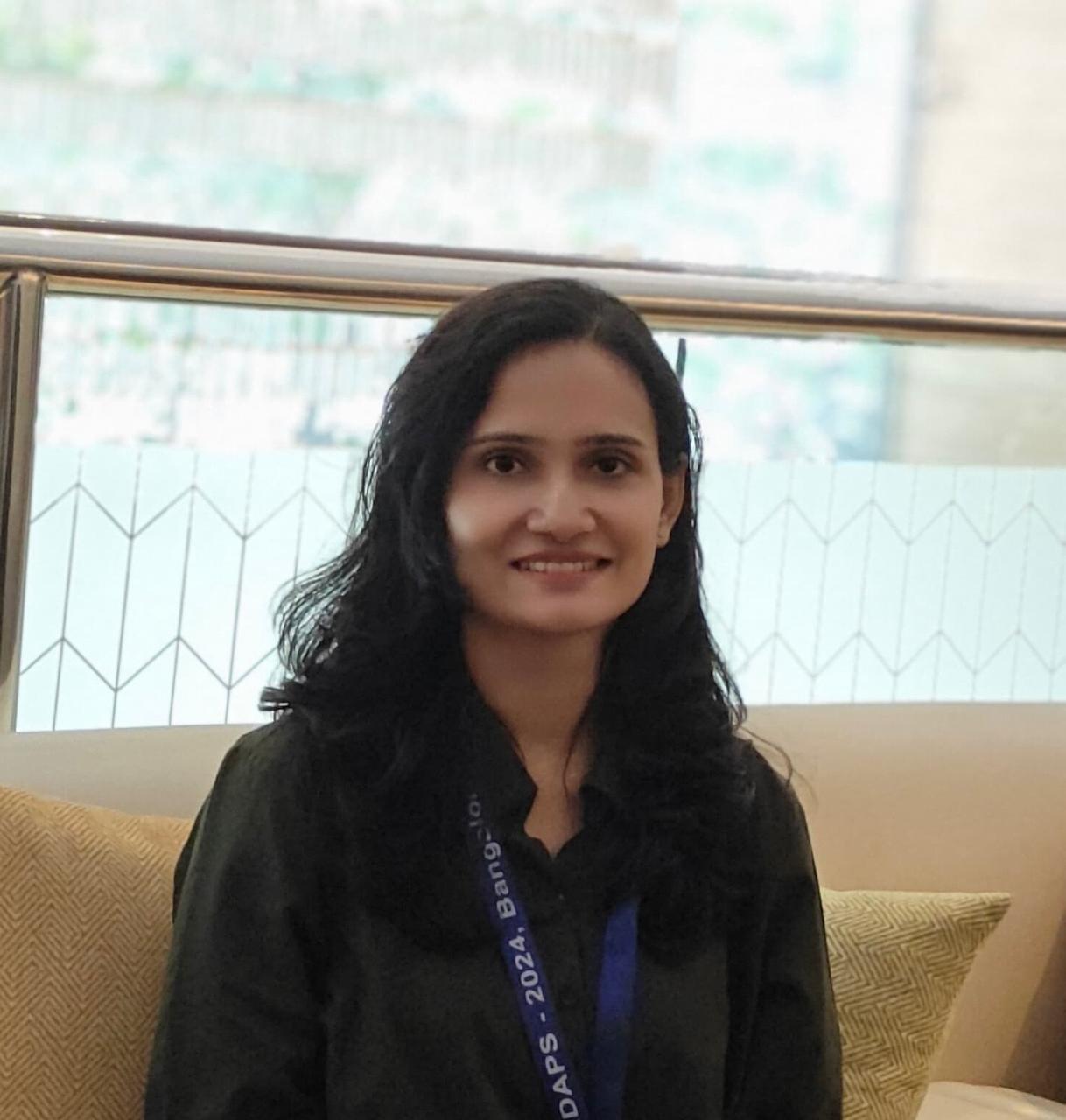}}]{Shilpa T N} (S'25) received BTech degree in Electronics and Instrumentation Engineering from Cochin University of Science and Technology and MTech in Micro and Nanoelectronics from College of Engineering Trivandrum. Since Aug 2021, She is pursuing PhD from the Department of Electrical Engineering, National Institute of Technology Rourkela.
\end{IEEEbiography}

\pagebreak

\begin{IEEEbiography}[{\includegraphics[width=1in,height=1.1in, clip,keepaspectratio]{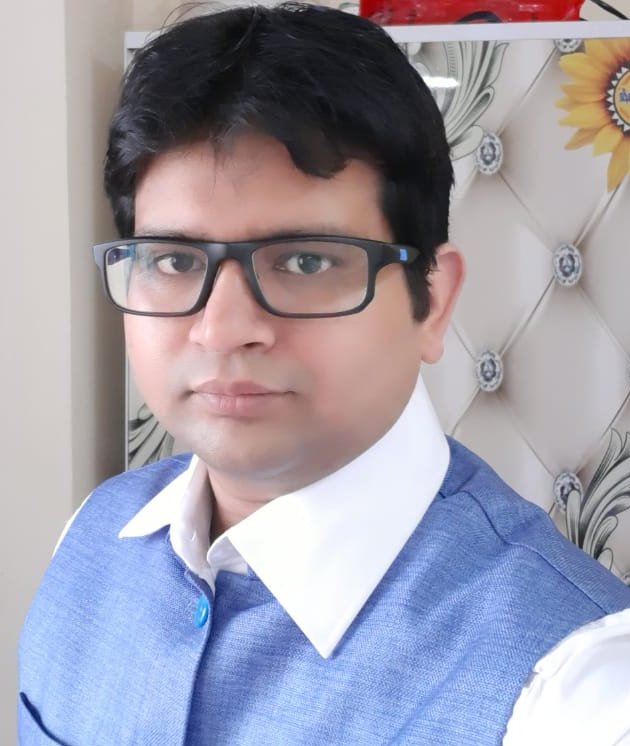}}]{Rakesh Sinha} (M'17)  received the B.Tech degree in electronics and communication engineering from Kalyani Government Engineering College, Kalyani, India, in 2008, and the M.Tech and the Ph.D. degree in RF and microwave engineering from the Indian Institute of Technology, Kharagpur, India, in 2011 and 2016 respectively.

From 2008 to 2009, he was with the Department of Robotics and Automation, Central Mechanical Engineering Research Institute, Durgapur, India, as a Junior Research Fellow. He was an Assistant Professor at the Department of Electronics \& Communication Engineering, JIS College of Engineering, Kalyani, India, from 2016 to 2017. During 2017-2018, he was associated with the Ulsan National Institute of Science and Technology, Ulsan, South Korea, as a Postdoctoral Researcher. He was associated with Chungnam National University, Daejeon, South Korea, from 2018 to 2019. Currently, Dr. Sinha is working as an Assistant Professor at the Department of Electrical Engineering, National Institute of Technology Rourkela, India. 

 His current research interests are in the area of multiport network synthesis, impedance matching, coupling, decoupling networks, phased array, and computational electromagnetic. Dr. Sinha has proposed the concept of the phase-shifting matching network, port-decomposition technique, theory of coupled characteristic mode, and the Y-matrix algorithm for impedance-transforming multiports. He and his students developed educational software for microwave circuit designs, which are available at IEEE Dataport and Zenodo.  Two undergraduate students, under the supervision of Prof Rakesh Sinha, received   Undergraduate Research Scholarships from  IEEE APS and MTTS, respectively.   
\end{IEEEbiography}
\end{document}

\title{Supplementary document for Fully-Adaptive and Semi-Adaptive Frequency Sweep Algorithm Exploiting Loewner-State Model for EM Simulation of  Multiport Systems}
\author{Shilpa T N and Rakesh~Sinha}
\date{}
\maketitle
\vspace*{1em}
\section{Additional Results}
In this section, additional results are provided to validate the proposed semi-adaptive frequency sweep method.
\subsection{Example 1 - Substrate Integrated Waveguide (SIW) microstrip transmission line transition}
\begin{figure}[!h]
\vspace{-1em}
\centering
   \includegraphics[width= 10 cm]{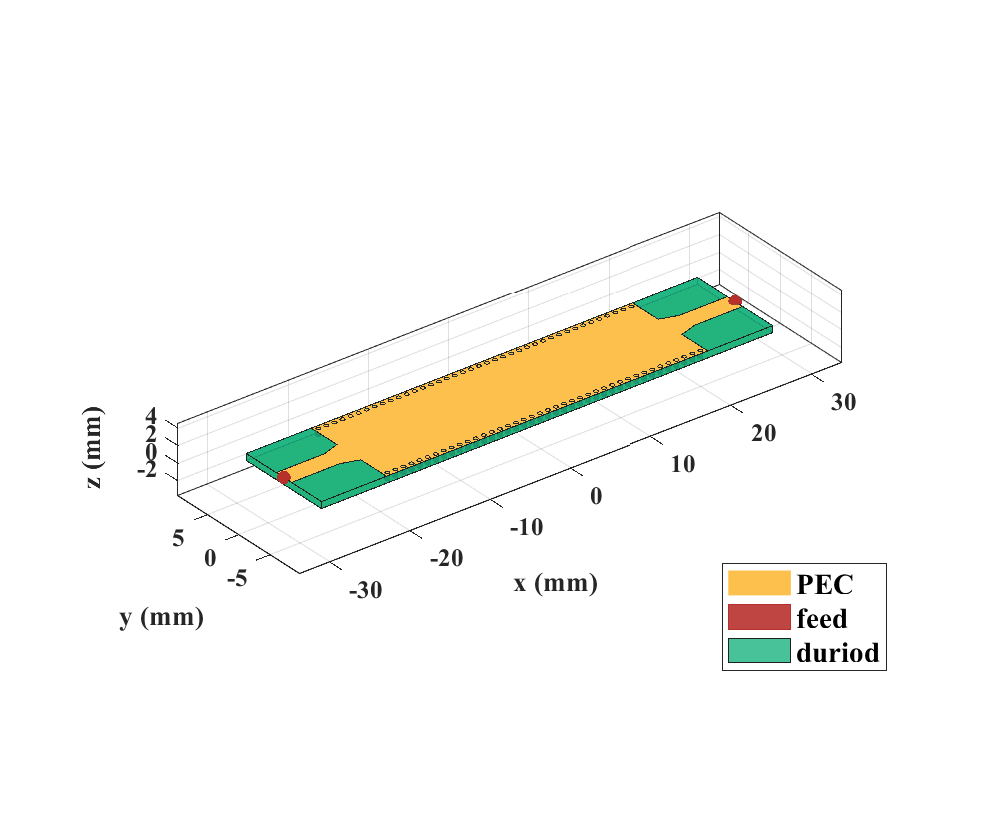}
\caption{\small  SIW Microstrip Transmission Line transition.}
\label{fig:SIW}
\end{figure}

\begin{figure}
    \centering
    % \subfloat[SIW]{\includegraphics[width= 9 cm]{figures/siw_pcb.eps}\label{fig:SIW}}
    \subfloat[Magnitude]{\includegraphics[width= 9 cm]{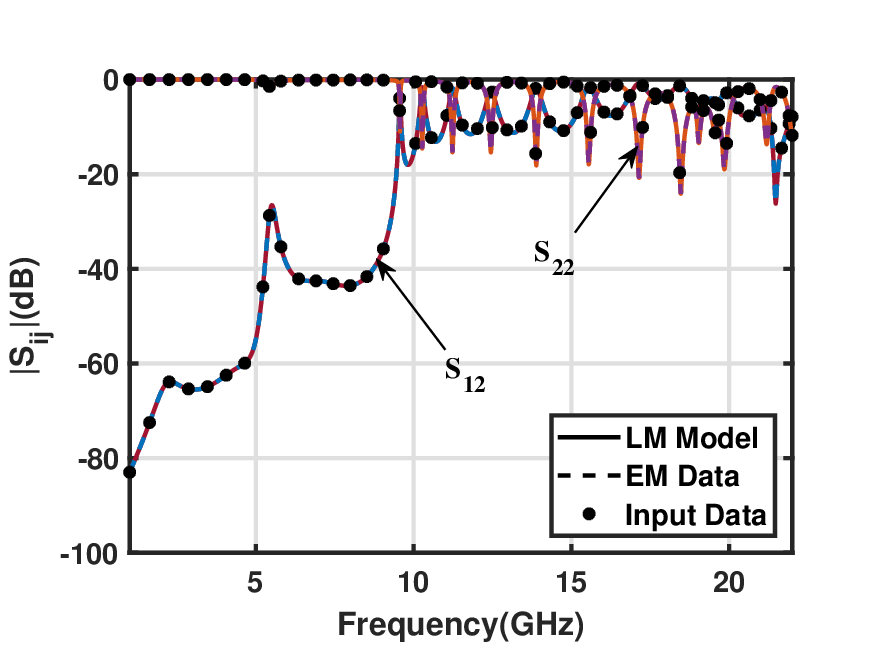}}\\
    \subfloat[Phase]{\includegraphics[width= 9 cm]{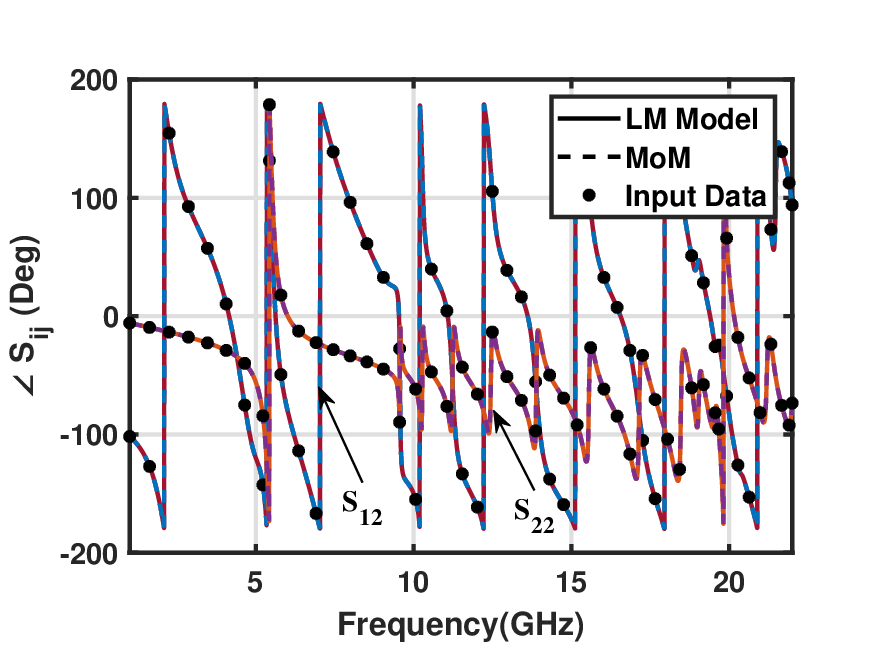}}
     \subfloat[Error]{\includegraphics[width= 9 cm]{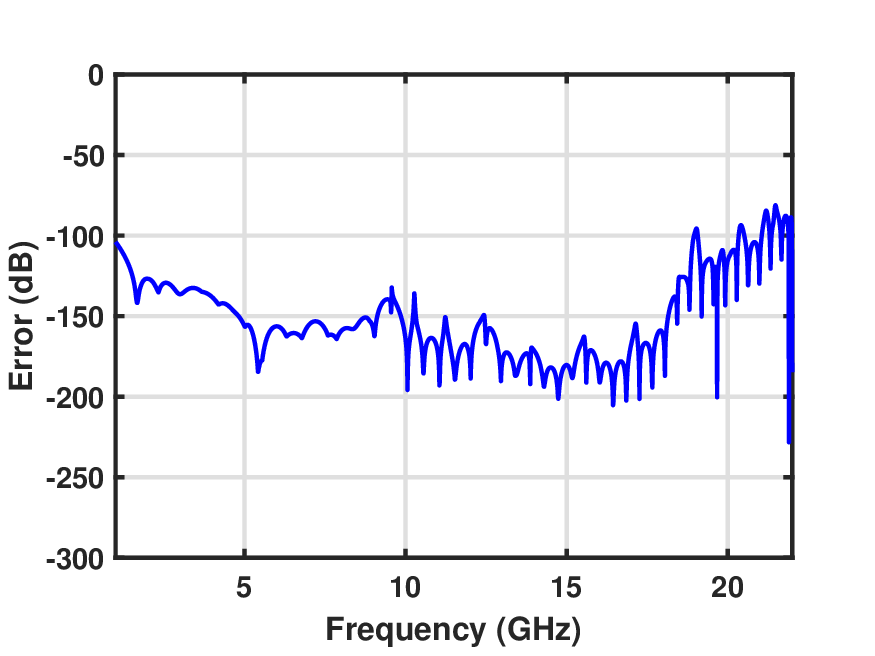}}
     \caption{Modeling the S-parameter data for SIW Microstrip Transmission Line.}
    \label{fig:siw2}
\end{figure}

This example considers a Substrate Integrated Waveguide (SIW) microstrip transmission line transition \cite{siw}, shown in Fig. \ref{fig:SIW}. This SIW tapered transmission line model is designed to operate with a 10 GHz to 15 GHz microwave frequency band. The substrate is 0.8 mm thick with $\varepsilon_\mathrm{r}=2.2 $, chosen as lossless. 
This particular case requires a frequency response in the range of 1 GHz to 22 GHz. The total length of the SIW microstrip transition is 56.2 mm. According to the third-order approximation, 46 frequency points are needed initially, for the semi-adaptive LM method. The time taken to generate the S-parameter data with a 10 MHz interval is  2114.32 seconds. The detailed simulation results are shown in Table \ref{tab:siw}. The magnitude, phase and error plots of the structure are shown in Fig. \ref{fig:siw2}.

\begin{table}[h]
    \centering
    \caption{Simulation Results of SIW Microstrip Transmission Line}
    \label{tab:siw}
\begin{tabular}{ |p{3cm}||p{3cm}|  }
\hline
\hline 
Method &  Semi Adaptive LM   \\
\hline
Input samples ($n$)  &  49 \\ 
\hline
 $t_{adpt.}$ (s) &  205.67 \\  
 \hline 
  $E_{max.}$ (dB) & -81 \\
  
  \hline 
 \end{tabular}
\end{table}

\subsection{Example 2 - Coplanar Waveguide}
\begin{figure}
    \centering
    \subfloat[CPW]{\includegraphics[width= 9 cm]{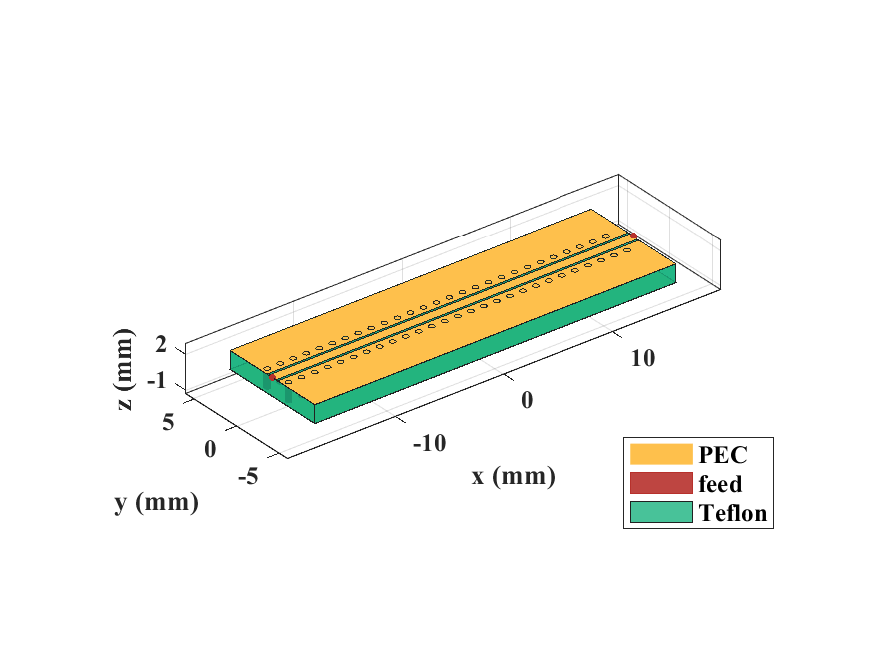}\label{fig:CPW}}
    \subfloat[Magnitude]{\includegraphics[width= 9 cm]{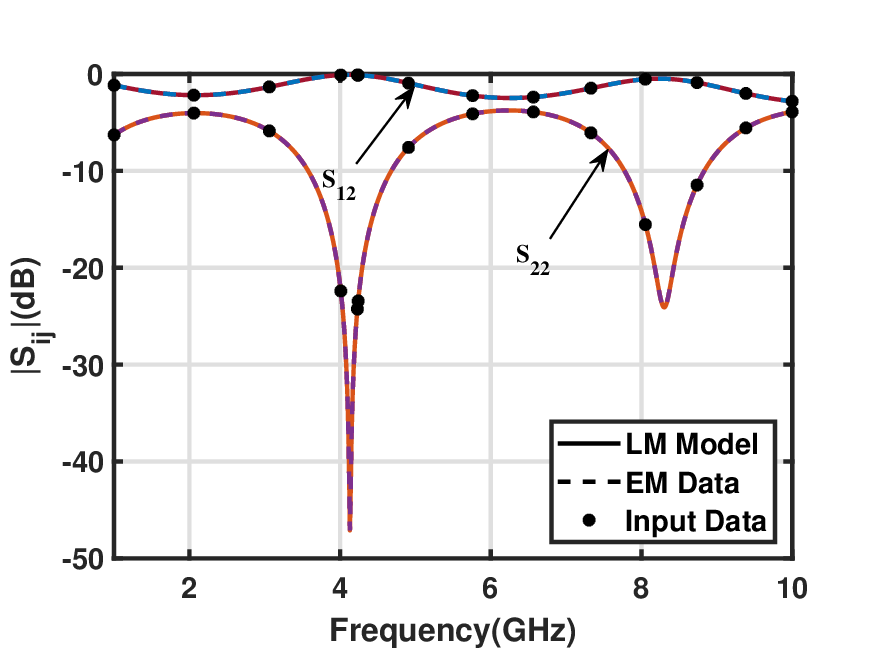}}\\
    \subfloat[Phase]{\includegraphics[width= 9 cm]{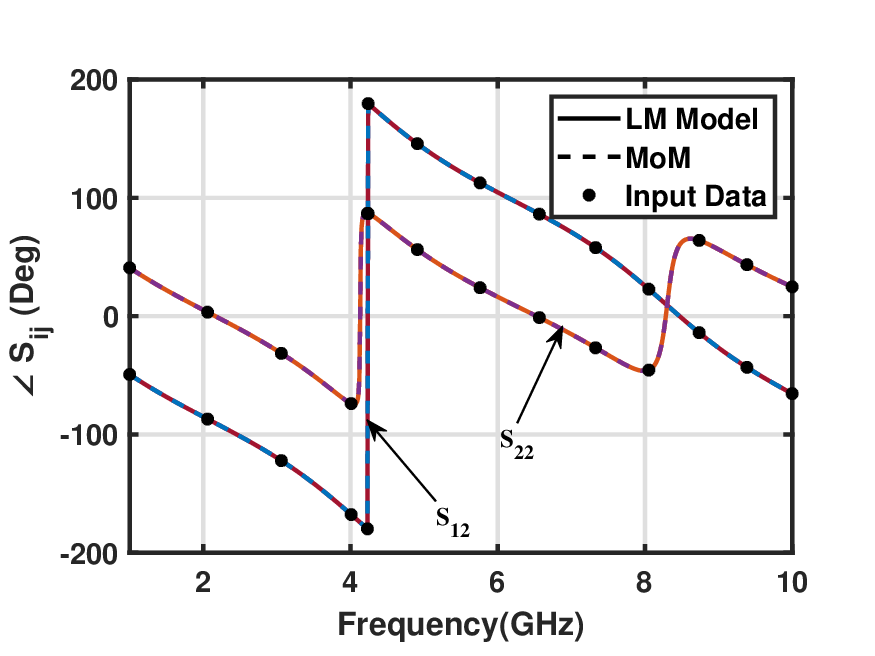}}
     \subfloat[Error]{\includegraphics[width= 9 cm]{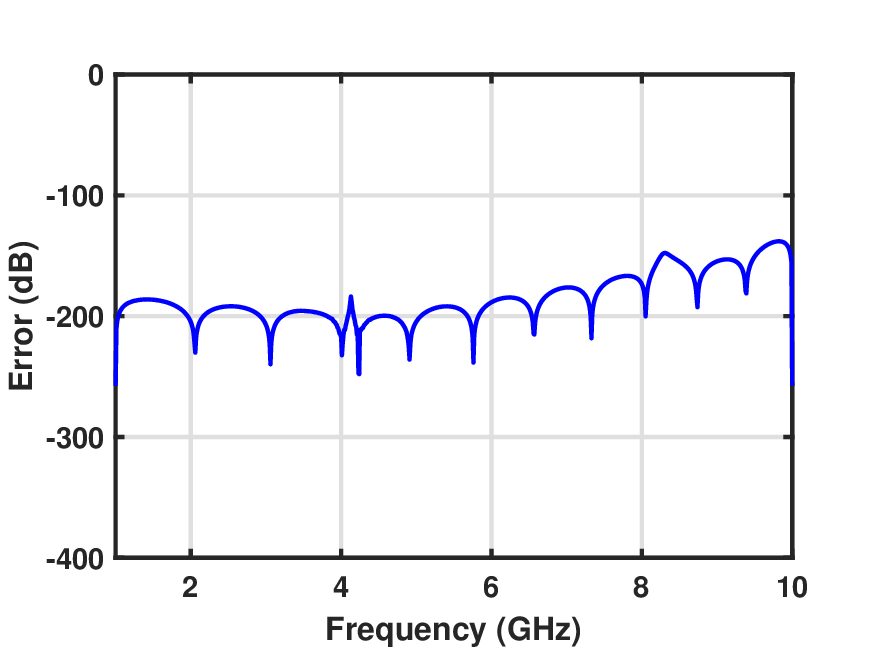}}
     \caption{Modeling the S-parameter data for Coplanar Waveguide.}
    \label{fig:cpw2}
\end{figure}

This example consists of a coplanar waveguide \cite{CPW}, using a Teflon substrate ($\epsilon_r=2.2$) as dielectric, as shown in Fig. \ref{fig:CPW}. The design frequency of the waveguide is 3 GHz, line length of 0.5 meters, and an impedance of $75\;\Omega$. The height of the substrate is 1.6 mm. This example requires a frequency response from 1 GHz to 10 GHz. The length of the waveguide is 33.3 mm. According to the third-order approximation, 12 frequency points are required initially for the semi-adaptive LM method. The time taken to generate the S-parameter data with a 10 MHz interval is 1271.21 seconds. The detailed simulation results are shown in Table \ref{tab:cpw}. The magnitude, phase and error plots of the structure are shown in Fig. \ref{fig:cpw2}.

\begin{table}[!h]
    \centering
    \caption{Simulation Results of Coplanar Waveguide.}
    \label{tab:cpw}
\begin{tabular}{ |p{3cm}||p{3cm}|  }
\hline
\hline 
Method &  Semi Adaptive LM   \\
\hline
Input samples ($n$)  &  14 \\ 
\hline
 $t_{adpt.}$ (s) &  23.45 \\  
 \hline 
  $E_{max.}$ (dB) & -138 \\
  
  \hline 
 \end{tabular}
\end{table}

\subsection{Example 3 - Stripline}
\begin{figure}[!t]
    \centering
    \subfloat[PCB]{\includegraphics[width= 9 cm]{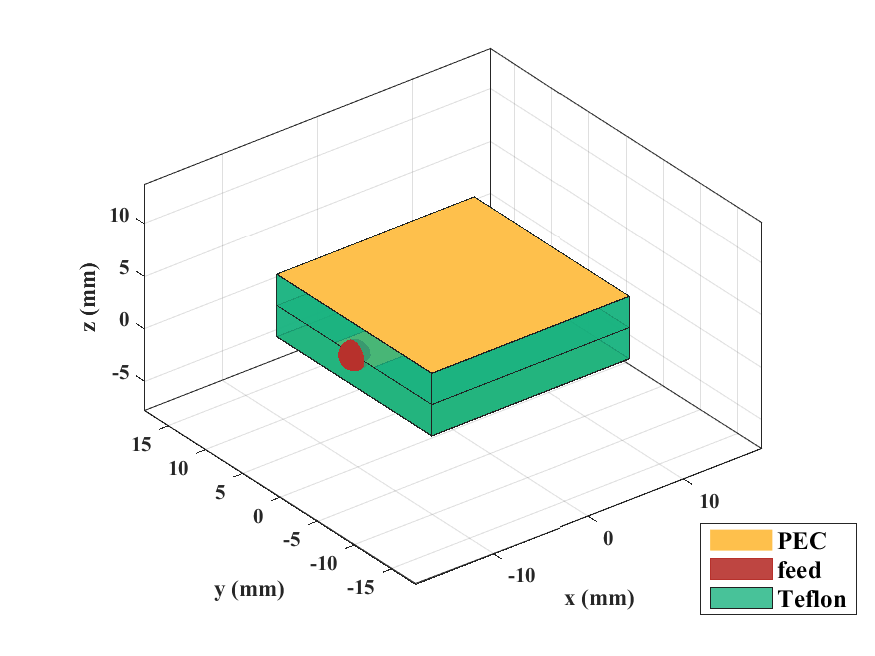}\label{fig:stripline}}
    \subfloat[Magnitude]{\includegraphics[width= 9 cm]{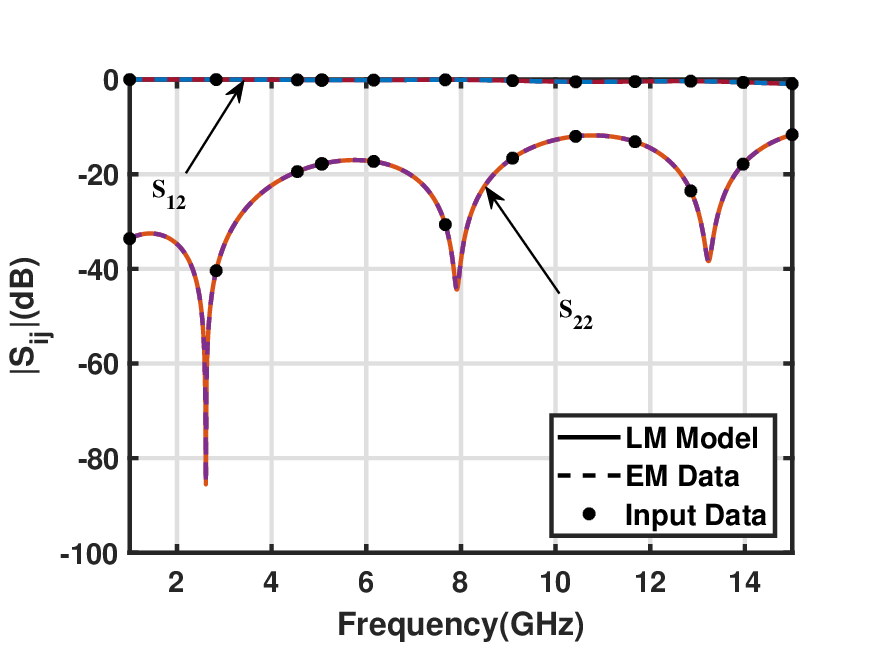}}\\
    \subfloat[Phase]{\includegraphics[width= 9 cm]{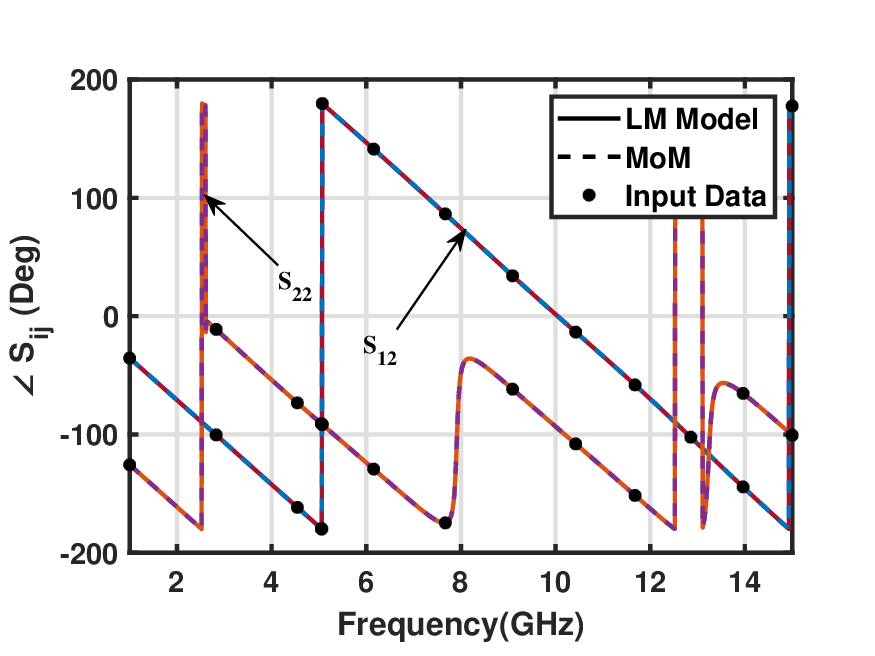}}
     \subfloat[Error]{\includegraphics[width= 9 cm]{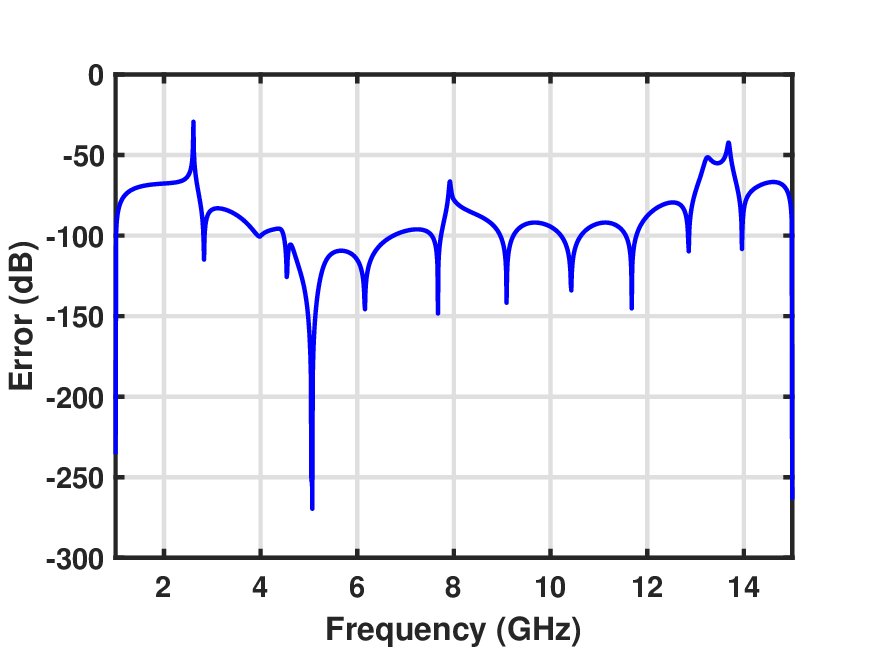}}
     \caption{Modeling the S-parameter data for Stripline.}
    \label{fig:stripline2}
\end{figure}

In this example, we have considered a stripline \cite{stripline}, as shown in Fig. \ref{fig:stripline}. Teflon substrate ($\epsilon_r=2.2$) has been used with a substrate height of h = 1.58 mm.  The total length of the line is 20.7 mm. According to the third-order approximation, 11 frequency points are needed initially for the semi-adaptive LM method. The time taken to generate the S-parameter data with a 10 MHz interval is  5084.98 seconds. The detailed simulation results are shown in Table \ref{tab:stripline}. The magnitude, phase and error plot of the structure is shown in Fig. \ref{fig:stripline2}.
% \begin{figure}[!h]
% \vspace{-1em}
% \centering
%    \includegraphics[width= 8 cm]{figures/stripline_normal_pcb.eps}
% \caption{\small  Stripline Transmission Line.}
% \label{fig:stripline}
% \end{figure}

\begin{table}[!h]
    \centering
    \caption{Simulation Results of Stripline.}
    \label{tab:stripline}
\begin{tabular}{ |p{3cm}||p{3cm}|  }
\hline
\hline 
Method &  Semi Adaptive LM   \\
\hline
Input samples ($n$)  &  14 \\ 
\hline
 $t_{adpt.}$ (s) &  40.84 \\  
 \hline 
  $E_{max.}$ (dB) & -29.34 \\
  
  \hline 
 \end{tabular}
\end{table}

\bibliographystyle{IEEEtran}
\bibliography{RREF}